\documentclass[aps,prb,longbibliography,floatfix,reprint]{revtex4-1}
\usepackage{amsmath}
\usepackage{graphicx}
\usepackage{amsfonts}
\usepackage{color}
\usepackage{placeins}

\begin{document}

\title{Phonon thermal conductivity by non-local non-equilibrium molecular dynamics}

\author{ Philip B. Allen }
\email{philip.allen@stonybrook.edu}
\affiliation{ Department of Physics and Astronomy,
              Stony Brook University, 
              Stony Brook, New York 11794-3800, USA }

\author{ Yerong Li }
\email{yerong.li@outlook.com}
\affiliation{ Department of Physics and Astronomy,
              Stony Brook University, 
              Stony Brook, New York 11794-3800, USA }
\affiliation {Department of Intensive Instruction, Nanjing University, Nanjing 210093, China}

\date{\today}

\begin{abstract}
 
Non-equilibrium (NE) molecular dynamics (MD), or NEMD,  gives a ``direct'' simulation of thermal conductivity $\kappa$.
Heat $H(x)$ is added and subtracted in equal amounts ($\int dx H(x)=0$) at different places $x$.  After
steady state is achieved, the temperature $T(x)$ is found by averaging over finite sections.
Usually the aim is to extract a value of $dT/dx$ from a place distant from sources and sinks of heat.
This yields a value $\kappa_{\rm eff}(L)$ for the thermal conductivity, $L$ being the system size.
The result is then studied as a function of $L$, to extract the bulk limit $\kappa$.  Here instead,
our heat is $H(x)=H_0 \sin(qx)$, where $q=2\pi /L$.  This causes a steady-state temperature $T_0 +
\Delta T \sin(2\pi x/L)$.  A thermal conductivity $\tilde{\kappa}(q)$ is extracted, which is
well converged at the chosen $q$ (or $L$).  Bulk conductivity $\kappa$ requires taking the $q\rightarrow 0$
limit.  The method is tested for liquid and crystalline argon.  One advantage is
reduced computational noise at a given total MD run time.  Another advantage is that $\tilde{\kappa}(q)$ has
a more physical meaning than $\kappa_{\rm eff}(L)$.  It can be easily studied using Peierls-Boltzmann
transport theory.  New formulas for $\tilde{\kappa}(q)$ in simplified Debye-type models give new insight
about extrapolation to $q\rightarrow 0$ or $1/L\rightarrow 0$.  In particular, it is shown that
$\kappa_{\rm eff}(L)$ is unlikely to behave as $\kappa -C/L$, and much more likely to behave
as $\kappa-C^\prime/\sqrt{L}$.  Convergence problems encountered in computational cells
with very large aspect ratios $L_\parallel/L_\perp$ are also analyzed.

\end{abstract}

\maketitle

\section{Introduction}

Molecular dynamics (MD) simulation applies to crystals provided the temperature
$k_B T$ is reasonably high on the scale of the lattice energies $\hbar \omega$.  Then
classical Newtonian trajectories give good thermal averages.  Going beyond harmonic lattice dynamics is
difficult with quantum mechanics, but easy with classical MD, not requiring perturbation theory.
A ``direct'' simulation of phonon thermal conductivity $\kappa$ by non-equilibrium molecular dynamics (NEMD)
was first done by Payton {\it et al.} \cite{Payton1967} in 1967, and is now common.  Good examples are
the comparative study by Schelling {\it et al.} \cite{Schelling2002} 
and the careful study of GaN by Zhou {\it et al.} \cite{Zhou2009}. 
A difficulty occurs because small $\omega_Q$ vibrational modes have long mean free paths $\Lambda_Q$,
but only modes with $\Lambda_Q < L$ (where $L$ is the length of the simulation cell) have their contribution 
to $\kappa$ correctly treated by NEMD.

Long mean free paths imply non-locality of $\kappa$.  Phonons passing through point $x$ bring
information about the temperature $T(x^\prime)$ over distances $|x-x^\prime|$ comparable
to their mean free path.  Non-locality
is widely acknowledged\cite{Sussmann1963,Levinson1980,Wilson1984,Happek1988,Mahan1988,Chen1996}, 
but not always considered a direct topic of study.   Recent interest in inhomogeneous situations with
boundary effects and spatially varying heat input $H(\vec{r})$
gives new impetus to non-local analysis
\cite{Siemens2010,Johnson2013,Regner2014,Maznev2011,Koh2014,Maassen2015,Turney2009}. 
Homogeneous systems are much simpler; non-locality is easy to include theoretically.  A non-local analysis 
has value that deserves recognition.

In NEMD simulation, 
heat $H(x)$ is added and subtracted in equal amounts ($\int dx H(x)=0$) at different places $x$.  After
steady state is achieved, the temperature $T(x)$ is found by averaging over finite discrete sections.
Usually the aim is to extract a value of $dT/dx$ from a place distant from sources and sinks of heat.
This yields a value $\kappa_{\rm eff}(L)$ for the thermal conductivity, $L$ being the system size.
The result is then studied as a function of $L$, to extract the bulk limit $\kappa=\kappa_{\rm eff}(L\rightarrow\infty)$. 
If heating $|H(x)|$ is weak, response is linear, and the
relation $\Delta T(x)=\int dx^\prime S(x-x^\prime)H(x^\prime)$ must hold.
In Fourier variables, this is $\tilde{\Delta} T(q)= \tilde{S}(q)\tilde{H}(q)$.  From energy conservation $dJ(x)/dx=H(x)$
and the non-local Fourier law\cite{footnote1} $J(x)=-\int dx^\prime \kappa(x-x^\prime)dT(x^\prime)/dx^\prime$, one
finds $\tilde{\kappa}(q)=1/q^2 \tilde{S}(q)$.  The ``tilde'' is used to indicate when a function is in reciprocal space
instead of direct space.  


In this paper, we show that
we can improve on $\kappa_{\rm eff}(L)$ by thinking non-locally.   The results of NEMD 
calculations can be considered to arise from $\tilde{\kappa}(q)$
for values $q=2\pi n/L$ (for periodically repeated slabs), or $q=\pi n/L$ (for terminated slabs).
Here $L$ is the length of the simulation cell.  The desired true bulk 
$\kappa = \tilde{\kappa}(q\rightarrow 0)=\int d\vec{r}\kappa(\vec{r})/V$, requires
extrapolation.  This extrapolation is best guided by theory aimed at $\tilde{\kappa}(q)$.
A useful strategy is therefore direct  computation of $\tilde{\kappa}(2\pi/L)$, using this to 
optimize the extrapolation to $L\rightarrow\infty$.  The reason why $\kappa=\tilde{\kappa}(0)$ is unavailable is
because $\tilde{H}(0)=\int d\vec{r}H(\vec{r})/V$ must vanish in order for a steady state to be allowed.

This paper does four things.  (1) We Fourier-analyze the
common version of NEMD where the system is periodic with length $L$ along the direction of heat flow,
and heat input and removal occurs in isolated slabs separated by $L/2$.
A rigorous relation between $\kappa_{\rm eff}(L)$ and $\tilde{\kappa}(q)$ is worked out.
(2) A method is given for direct MD simulation
of $\tilde{\kappa}(q)$, by applying and extracting heat in a sinusoidal pattern \cite{Allen2014}.  
This has numerical advantages over other
protocols.  (3) Convergence of $\tilde{\kappa}(q)$ towards $\kappa$
is studied by NEMD simulation
for the Lennard-Jones (LJ) liquid and crystal.  (4) The Peierls ``Boltzmann Transport Equation'' (BTE) 
\cite{Peierls1929,Ziman1960} is solved for $\tilde{\kappa}(q)$ in Debye approximation ($\omega_Q=v|\vec{Q}|$)
with $1/\tau_Q = (1/\tau_D) (\omega_Q/\omega_D)^p$.  The appropriate power $p$ is probably 2.
This helps understand convergence as the MD simulation-cell size $L$ increases.  For the particular case
$p=2$ most often encountered, it is shown that $\kappa_{\rm eff}(L)\sim\kappa-C^\prime/\sqrt L$, rather than
the form $\sim \kappa-C/L$ which has been widely assumed.

\section{Preliminaries}

Assume a homogeneous single crystal, represented by a 
simulation cell with periodic boundary conditions.  
The $i^{\rm th}$ atom at
$\vec{r}_i$ and the atom at $\vec{r}_i+\vec{R}$ are equivalent, and have the same 
trajectory $\vec{r}_i(t)$.  For simplicity, the 
primitive translation vectors $\vec{R}$ of the simulation cell ($\vec{A}_1, \   \vec{A}_2, \  \vec{A}_3$), are
assumed orthogonal.  For example, in the sample calculations presented later, they are taken
to be $N_x a \hat{x}, \ N_y a \hat{y}, \ N_z a \hat{z}$, where $a$
is the lattice constant, the edge-length of the {\it fcc} conventional cube.   A typical cell has size
$(N_x, \ N_y, \ N_z)=(80, \ 6, \ 6)$,  With 4 atoms in the conventional cube, this means
11,520 atoms.  Heating is done in slabs perpendicular to the long vector ${\vec A}_1$.
Therefore, current flows parallel to $\vec{A}_1$.  This is why a one-dimensional
notation ($J(x)$ and $\tilde{\kappa}(q)$, for example) is used.  The simulation cell length
$L=L_x= |\vec{A}_1|=N_x a$ is chosen as large as computation permits, trying to surpass the 
distance $\Lambda$ of non-local thermal memory.

Boundaries create challenging problems.  Nanoscale heat transfer is typically
dominated by thermal boundary (or Kapitza) resistance \cite{Swartz1989,Cahill2014}.
For study of bulk conductivity, the aim is to
reduce the influence of boundaries.  One can argue \cite{Liang2014} that periodic boundary 
conditions are not the most favorable way to do this.  However, in this paper, periodicity offers
simplicity for analysis, overruling other considerations.   

A further simplification follows computational necessity.  Discretize the temperature $T(\vec{r})$
into slab values $T(\ell)$.  The slabs are layered
in the $\vec{A}_1$-direction, and have width $d=n_S a$ where $n_S$ is a small
integer and a factor of $N_x$.   The number of slabs is $N_S = N_x/n_S$.
Let the variable $x$ denote distance along
the $\vec{A}_1$-axis, perpendicular to slabs.  The slab indexed by the integer $\ell$
occupies the interval $\ell d -d/2< x<\ell d + d/2$.  The temperature $T(\ell)$ is found from
the average kinetic energy of the atoms in the $\ell$'th slab.  Heat is transferred
externally into the $\ell$'th cell at a volume-average rate $H(\ell)$.  In steady state,
an outward heat flux conserves energy, $H(\ell)=[J(\ell+1/2)-J(\ell-1/2)]/d$.  Both current and
temperature gradient are properties of the junction of two adjacent slabs.  Their steady-state
linear relation is $J(\ell+1/2)=-\sum_{\ell^\prime}
\kappa(\ell,\ell^\prime)\nabla T(\ell^\prime+1/2)$, where the sum runs over the $N_S$ slabs.  
This definition is required by linear math.
Periodicity requires $\kappa(\ell,\ell^\prime)=\kappa(\ell+mN_S,\ell^\prime+nN_S)$,
and homogeneity (if the medium is in fact homogeneous) requires that $\kappa(\ell,\ell^\prime)=
\kappa(\ell-\ell^\prime)$.  Corresponding ({\it via} a unitary Fourier transformation) to the $N_S$ distinct slabs, 
there are $N_S$ distinct wave-vectors $q_n=2\pi n_q/N_S d=2\pi n_q/L$, indexed by integers
$n_q$ ($|n_S|\le N_S/2$), and distributed in a one-dimensional Brillouin zone.  In the homogeneous
case, the relation is $\tilde{J}(q)=-\tilde{\kappa}(q)\tilde{\nabla} T(q)$.  
These ideas were introduced in Ref. \onlinecite{Allen2014}, where further properties are explained.  

It is not hard to
extend the usual derivation of the Kubo formula (ref. \onlinecite{AllenFeldman1993}, for example)
to derive a Kubo formula for $\kappa(x,x^\prime)$ or $\kappa(\ell,\ell^\prime)$.  The
classical limit is
\begin{equation}
\kappa(\ell,\ell^\prime)=-\frac{1}{k_B T^2}\int_0^\infty dt < J(\ell+1/2,t)J(\ell^\prime+1/2,0)>
\label{eq:Kubo}
\end{equation}
%


\section{Discrete slab Heating}

A common version of NEMD simulation removes
energy only from slab $\ell=0$, at a volume-average rate $H$, 
and inserts energy at the same rate into slab $\ell=N_S/2$.
Zhou {\it et al.} \cite{Zhou2009} did a careful
study of $\kappa$ for GaN by this method.  They discuss, but do not completely resolve, the
issue of how the answer for $\kappa$ scales with system size.  Here we analyze this
version with periodic boundaries ($\ell=\ell+N_S$), rather than the rigid or free boundaries sometimes used. 
Heat current $J_x(\ell+1/2)=\pm (d/2)H \equiv \pm J$ flows across
slab boundaries, the plus sign for $\ell$ to the right of the input and left of the output slab, and
the minus sign for opposite cases.  Thermal conduction is found using the
relation  $\kappa_{\rm eff}(L) \equiv -J/\nabla_x T({\rm mid})$.  The rigorous non-local
Fourier law is $\tilde{J}(q_n)=-\tilde{\kappa}(q_n)\tilde{\nabla}T(q_n)$.  
Analysis given in Appendix \ref{App:keff} shows that 
\begin{equation}
\frac{1}{\kappa_{\rm eff}(L)} = \frac{2}{N_S}\sum_q^{n_q={\rm odd}} \frac{(-1)^{(n_q-1)/2}}{\sin(qd/2)} 
\left[ \frac{\cos(qd/2)}{\tilde{\kappa}(q)} \right] 
\label{eq:keffq}
\end{equation}
where the subscript on $q_n$ has been dropped.
This holds if $N_S/4$ is an integer.  If it is a half-integer, then $\cos(qd/2)$ is replaced by 1.
Liquids have a very local conductivity, where $\tilde{\kappa}(q)\approx\kappa$
is nearly independent of $q$.  Then the sum in Eq.(\ref{eq:keffq}) converges correctly
to $\kappa_{\rm eff}\rightarrow\kappa$ without needing a small $q_1=q_{\rm min}=2\pi/L$.
For crystals like GaN, where non-local behavior is caused by long mean free paths of small-$Q$
phonons, $\tilde{\kappa}(q)$ peaks around $|q|=0$.  Accurate results
require a small $q_{\rm min}$ or a large $L$.  When $q_{\rm min}$ is not
very small, the first two terms $n_q=\pm 1$ in Eq.(\ref{eq:keffq}) dominate.  Approximating
$\sin(q_{\rm min}d/2)$ by $\pi/N_S$ and $\cos(q_{\rm min}d/2)$ by 1, the result is
$\kappa_{\rm eff}\approx \pi\tilde{\kappa}(q_{\rm min})/4$, which is smaller than $\pi\kappa/4$
(and still smaller than the true $\kappa$.)  It would be better to calculate
$\tilde{\kappa}(q_{\rm min})$ directly.  A method is given in the next section.

\par
\begin{figure}
\includegraphics[angle=0,width=0.5\textwidth]{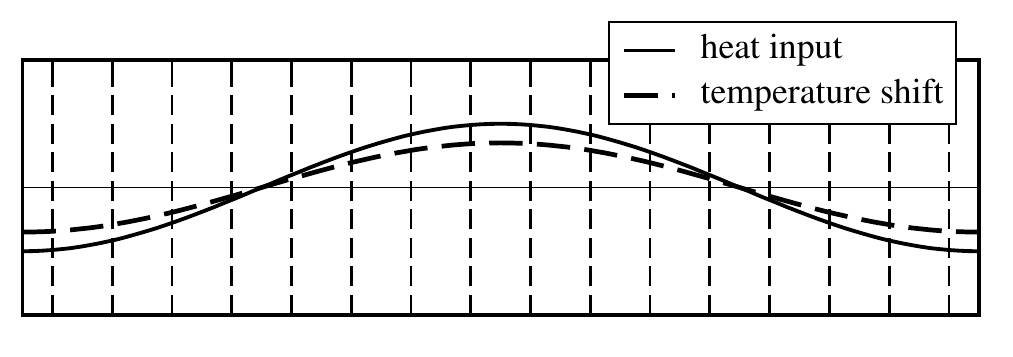}
\caption{\label{fig:slab} Schematic picture of an $N_S=16$ slab simulation cell
with periodic boundary conditions.  Heat is
added or subtracted according to position, in the $\ell^{\rm th}$ slab, as $H\cos(2\pi \ell/N_S)$.
The central cell is numbered $\ell=0$.
Temperature $T(\ell)$ is determined by averaging kinetic energy of atoms in each slab.  In linear
approximation, the steady-state temperature must vary as $T_0+\Delta T\cos(2\pi \ell/N_S)$.}
\end{figure}
\par

\section{Sinusoidal Heating Algorithm}
\label{sec:Sine}

The simulation cell is divided into slabs centered at $x(\ell)=\ell d, \  \ell=1,\ldots,N_S$.  It
is shown schematically in Fig. \ref{fig:slab}.   The distance $d=L/N_S$ is the width of a
slab.  It is a multiple of $a=L/N_x$ where $N_xa=L$ is the repeat distance on the long axis.
We want to modify
the heat input profile.  Instead of the two slab version, let the heat input be of the form
$H(\ell)=\tilde{H}(q_{\rm min})\cos(q_{\rm min}x(\ell))$.  
The temperature variation then has the form $T(\ell)=T_0+\tilde{\Delta} T(q_{\rm min}) \cos(q_{\rm min}x(\ell))$. 
Since slab temperatures $T(\ell)$ at all of the $N_S$ different values of $\ell$ 
are used to compute the Fourier amplitude $\tilde{\Delta} T(q_{\rm min})$, numerical noise 
averages out faster.

Equation 26 of the previous paper \cite{Allen2014} says, for arbitrary
heating $H(\ell)$, heating rate, temperature, and conductivity
in Fourier space are related by (for $q\ne 0$), 
\begin{equation}
\tilde{\Delta} T(q)=\frac{\tilde{H}(q)d^2}{4\sin^2 (qd/2) \tilde{\kappa}(q)}
\label{eq:nonlocal}
\end{equation}
For simple sinusoidal heating, $\tilde{H}(q)$ and $\tilde{\Delta} T(q)$ are zero except
at $q=q_{\rm min}$, where their values are denoted as $H$ and $\Delta T$.  Then the thermal conductivity is
\begin{equation}
\tilde{\kappa}(q_{\rm min})=\frac{Hd^2/\Delta T}{4\sin^2(q_{\rm min} d/2)}
\rightarrow \left(\frac{N_S d}{2\pi}\right)^2 \frac{H}{\Delta T}
\label{eq:kappa}
\end{equation}
This makes sense:  $2\pi\Delta T/N_S d$ is the maximum temperature gradient, 
and $HN_S d/2\pi$ is the maximum heat current.

Now we need a good numerical algorithm to drive the oscillatory
heat input.    Furtado, Abreu, and Tavares\cite{Furtado2015} (FAT) made an improvement on the
popular algorithm by M\"uller-Plathe \cite{Muller1997}.  The usual M\"uller-Plathe
method gives equal heating and cooling to two chosen slabs.  The hottest atom in the slab chosen for heat removal,
and the coldest atom in the slab for heat insertion, have their velocities interchanged,
conserving energy and momentum.  The FAT algorithm does not interchange velocities.
It is decided in advance what heat $\Delta \epsilon$ should be added and subtracted.  
Then a corresponding velocity increment 
$\Delta \vec{v}$ is added to one of the two atoms, and subtracted from the other, 
in such a way that total energy and momentum are conserved, local energy being altered by $\pm\Delta\epsilon$.  
The minimum possible magnitude $|\Delta \vec{v}|$, is chosen, so that the resulting disruption 
is minimized.  This allows heat input at a
predetermined rate which can be spatially varied.  
Details are given in the on-line supplemental material\cite{footnote2}.


\section{Test on Lennard-Jones liquid}
\label{sec:LJL}

The Lennard-Jones (LJ) liquid is a simple case, used by M\"uller-Plathe \cite{Muller1997} to test his algorithm.
The pair potential is
\begin{equation}
V_{\rm LJ}=4\epsilon \left[ \left (\frac{\sigma}{r}\right)^{12} - \left (\frac{\sigma}{r}\right)^6 \right]
\label{eq:VLJ}
\end{equation}
The parameters for argon \cite{Michels1949} are $\epsilon/k_B=119.6$K
and $\sigma$ =3.405\AA. 
First, we use M\"uller-Plathe's algorithm to reproduce his results, at the same $(N,V,T)$=
2592 atoms, $\rho=N/V=0.849/\sigma^3$, and $T=0.7\epsilon/k_B=$84K. The same simulation
cell is used, consisting of $18\times 6\times 6$ {\it fcc} conventional cubes, and
periodic boundary conditions.
The cut-off distance for the LJ potential is $3.0\sigma$.  We get the same 
answer, $\kappa$=7.1 in LJ units.

\begin{figure}
\includegraphics[angle=0,width=0.5\textwidth]{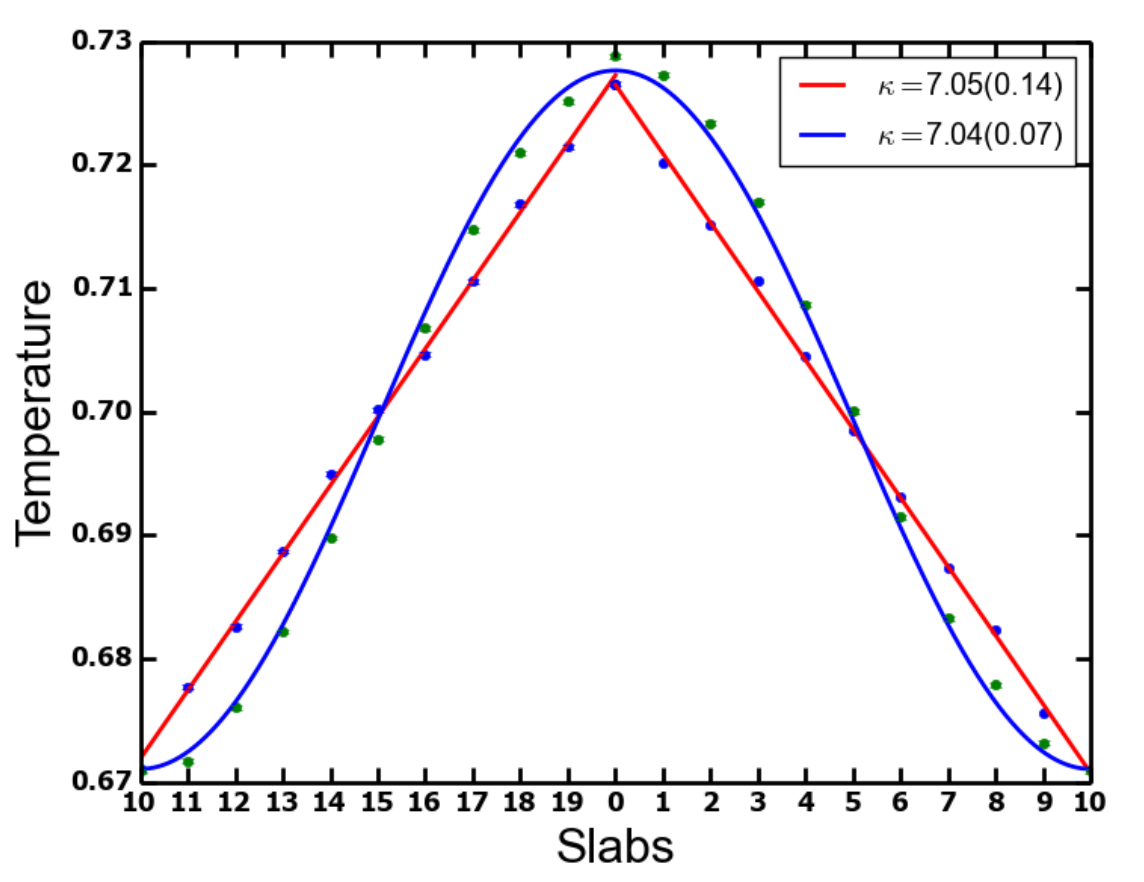}
\caption{\label{fig:amplitude} Temperature profiles from NEMD simulations for the 
LJ liquid.  The points fit by the straight line are a reproduction of the
calculation of M\"uller-Plathe.  The points fit by the sine curve use the same simulation
cell, with spatially sinusoidal heating and cooling.}
\end{figure}

As shown in Fig. 2 of ref. \onlinecite{Muller1997},
and confirmed by our calculation in Fig. \ref{fig:amplitude}, 
the temperature gradient is essentially constant all the way to, and including, 
the slabs $0$ and $N_S/2=10$.  This is because thermal conductivity in a liquid is
very local.  This can be contrasted with Fig. 2 of ref. \onlinecite{Zhou2009} or Fig. 4
of ref. \onlinecite{Allen2014}, for crystals with non-local $\kappa$.
Gas theory is certainly not correct for a liquid; the concept of a mean-free
path is not valid.  However, we can get an idea of what happens by unlicensed use of the gas 
formula $\kappa = C{\bar v}\Lambda/3$.  The measured thermal conductivity
of liquid argon (0.132 W/mK at temperature near 100K and pressure 
near 1Mbar \cite{Younglove1986, Cook1961, Hanley1974, Ziebland1958})
then corresponds to a mean free path $\Lambda \approx 0.14 \ \AA$, more
than 30 times smaller than the slab separation $a=1.68\sigma$.  In other words, the non-local
conductivity $\kappa(z-z^\prime)$ decays to zero by the first neighbor slab, or
$\tilde{\kappa}(q)$ is independent of $q$ out to values of $q$ larger than $q_{\rm max}=\pi/d$. 

\par
\par

Also shown in Fig. \ref{fig:amplitude} is the sinusoidal temperature profile gotten
numerically from our sinusoidal heating.  The computational system is unaltered.  The  
2592 atoms are in the same cell, divided into 20 slabs, at the same $T$.    Heat $H(\ell)=H
\cos(2\pi\ell/N_S)$ was inserted, with $H$ between 1 and 5 $\epsilon/(\Delta t\cdot {\rm slab})$.  
Slabs were heated by choosing 
``adjoint" slabs ($\ell$ and $\ell+N_S/2$), finding coolest and hottest atoms, and 
altering the kinetic energies sinusoidally by use of the FAT algorithm. 
This was done for all slabs simultaneously, at a fixed time step 
($\Delta t=15\delta t$ for short samples and $60\delta t$  for long samples).
The time step $\delta t$ used for the ``velocity Verlet" Newton's-law integration algorithm
\cite{Hansen1969, Tuckerman1991} is $\delta t = 0.007 t_{LJ}$.  The
LJ unit of time for argon is $t_{LJ}=\sigma\sqrt(m/\epsilon)=2.16$ ps.
Equilibration required $10^4 \ \delta t$ of constant $T$ simulation, and $T(\ell)$ averaging was done
for $2\times10^5 \delta t$; good convergence was found in $5\times10^4 \delta t$ as shown in Fig.
\ref{fig:error}.

To estimate errors, consider that there are 130 atoms per slab, each with mean energy $k_B T$
and rms deviation of $k_B T$ from the mean, according to Maxwell-Boltzmann statistics.  
Thus the mean slab energy per atom, at any particular moment,
should be about $k_B T \pm k_B T/\sqrt 130$  Therefore, if averaged over 100 random and independent
thermalized configurations, the temperature error in a slab will be less than 1\%.  A run of
$5\times10^4 \delta t$ should be more than sufficient for this purpose.  Fig. \ref{fig:amplitude}
suggests errors of order 0.001$k_B T$ in the slab temperatures.

Both of the current LJ liquid simulations give the same
value, $\kappa$ = 7.1 in LJ units, equal to the M\"uller-Plathe \cite{Muller1997} result.
In normal units, this is  0.133 W/mK, very close to the experimental
value for argon, 0.132 W/mK \cite{Younglove1986,Cook1961,Hanley1974,Ziebland1958}.  
The sinusoidal algorithm gives
faster convergence and a slightly more accurate final answer, as shown in
Fig \ref{fig:error}.

\par
\begin{figure}
\includegraphics[angle=0,width=0.5\textwidth]{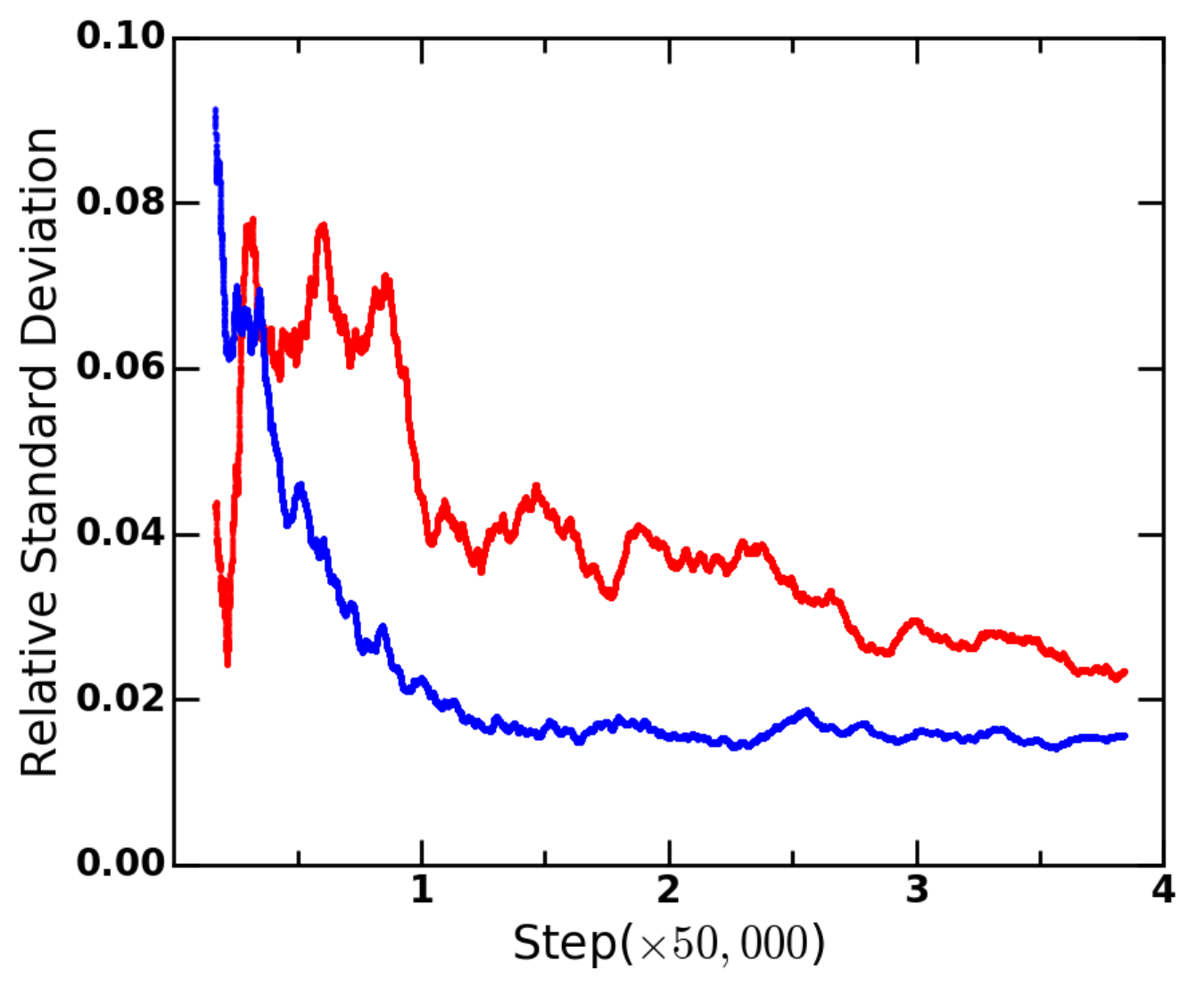}
\caption{\label{fig:error} Time evolution of the error of simulation of $\kappa$ for the
LJ liquid.  The upper and noisier curve uses 
The lower curve is the result of the sinusoidal algorithm of this paper.}
\label{fig:error}
\end{figure}
\par


\section{Extrapolation}
\label{sec:extrap}

NEMD answers for $\kappa$ are computed for finite size $L$.  Therefore extrapolation is required
to estimate the bulk ($L\rightarrow \infty$) answer.  Sellan {\it et al.} \cite{Sellan2010} have analyzed this.
It was also analyzed in the previous paper \cite{Allen2014}, using a Debye model.  
Here we continue the analysis.
Equations (22,23) of ref. \onlinecite{Allen2014} are
\begin{equation}
\kappa(q)=\frac{1}{\Omega}\sum_Q \hbar\omega_Q\frac{\partial n_Q}{\partial T} 
v_{Qx}^2 \tau_Q \cos^2(qd/2)   F(q,\Lambda_{Qx})
\label{eq:kappaslab1}
\end{equation}
\begin{equation}
F(q,\Lambda)=  \left[ 1 +4\sin^2 (qd/2)
\left\{ \left( \frac{\Lambda}{d} \right) + \left( \frac{\Lambda}{d} \right)^2 \right\} \right]^{-1} ,
\label{eq:kappaslab2}
\end{equation}
where $v_{Qx}, \ \tau_Q$, and $\Lambda_{Qx}=v_{Qx}\tau_Q$ are the group velocity, quasiparticle lifetime,
and $x$-component of
mean free path of the phonon mode of frequency $\omega_Q$.   The sum over modes $Q$ implicitly
includes a sum over branches.  These equations solve the BTE
in relaxation-time approximation (RTA; also known as the ``Single Mode Approximation'') 
for the case where heat is applied as $H(\ell)\propto\cos(q\ell d)$.  We focus on the smallest
$q$, $2\pi/L$.

Recent progress in numerical solution of the BTE \cite{Omini1995, Broido2007, Li2014, Fugallo2013} 
has enabled comparison of RTA against exact solutions.  Very often,
RTA answers are accurate at room temperature, 
graphene \cite{Lindsay2010, Fugallo2014} being a notable
exception.   Here we adopt both the RTA and the overly simplistic Debye model.  The aim is not 
an accurate $\kappa$, only insight about the size-dependence of $\kappa(q_{\rm min}=2\pi/L)$, to
guide extrapolation to the $L \rightarrow\infty$ limit.  We choose the mean free path $\Lambda_{Qx}$
to scale as $Q^{-p}$,
\begin{equation}
\Lambda_{Qx}=v\tau_D \frac{Q_x}{Q} \left(\frac{Q_D}{Q}\right)^p.
\label{eq:D3}
\end{equation}
Then Eqns.(\ref{eq:kappaslab1}, \ref{eq:kappaslab2}) become
\begin{equation}
\kappa_D(q)=\kappa_{0} \cos^2 (qd/2) \frac{1}{N}\sum_Q^{3N} \left( \frac{Q_x}{Q}\right)^2
\left(\frac{Q_D}{Q}\right)^p F(q,\Lambda_{Qx}).
\label{eq:D1}
\end{equation}
Here the sum over $Q$ contains an explicit factor of 3, for the three acoustic branches,
all given the same velocity $v$ in the Debye model; 
$\kappa_{0}$ is a convenient scale factor, 
\begin{equation}
\kappa_{0}=\frac{N}{\Omega} k_B v^2 \tau_D.
\label{eq:D2}
\end{equation}
This is just the classical limit
of the standard formula $(1/3)Cv\ell$ with $\ell=v\tau_D$.
The Debye wavevector has its usual value, $(6\pi^2 N/\Omega)^{1/3}$.  A 
common choice for the exponent is $p=2$.

Here, instead of integrating over the Debye sphere, we use direct numerics to do the 
discrete sum of Eq. \ref{eq:D1} over the actual discrete $Q$-mesh in the 
Brillouin zone of the {\it fcc} simulation cell
that will be used in the next section for the LJ crystal.  That is, we use only those
$\vec{Q}$'s  in the {\it fcc} Brillouin zone such that $\exp(i\vec{Q}\cdot\vec{A}_i)=1$,
where $\vec{A}_i$, for $i=1,2,3$, are the orthogonal translation vectors of the simulation
supercell.  As an example, the mesh used in Sec. \ref{sec:LJL} for the LJ liquid corresponds
to $\vec{A}_1 = 18a(1,0,0)$, $\vec{A}_2 = 6a(0,1,0)$, and $\vec{A}_3 = 6a(0,0,1)$,
where $a=1.68\sigma$ is the lattice constant of the {\it fcc} conventional cubic cell, using the
liquid argon density, 0.849/$\sigma^3$.
Our simulation cells in this and the next section will be very similar, but longer in the $\vec{A}_1$ direction,
{and with $a$ readjusted to $1.56\sigma$ to give the higher density\cite{Tuckerman1991,Ladd1978},
1.053/$\sigma^3$, of the low pressure LJ crystal.
The corresponding $\vec{Q}$'s are the vectors $\ell\vec{G}_1+m\vec{G}_2+n\vec{G}_3$
of the lattice reciprocal to the $\vec{A}$'s.  This
 is an anisotropic reciprocal-space mesh, being coarse in the directions $\vec{A}_2$ and $\vec{A}_3$, but finer
in the direction $\vec{A}_1$, corresponding to the actual distribution of normal modes of the atoms
in the simulation cell of the LJ crystal (where $|\vec{G}_{2,3}|$ is larger than $|\vec{G}_1|$.) 
We are guessing that the Debye model, with frequency
$\omega_Q = v|\vec{Q}|$ for all 3 branches, and $1/\tau_Q = (1/\tau_D)(Q/Q_D)^2$, sufficiently
captures the physics of the LJ crystal for purposes of learning how to extrapolate to infinite
simulation cell size. 

Results are shown in Fig. \ref{fig:divergence15} and in Appendices \ref{app:B} and \ref{app:C}. 
The figure shows two things.  First, quite smooth extrapolation to the correct $q=0$ answer 
appears when $\kappa(q)$ is plotted {\it versus} $\sqrt{qa}$, as anticipated in Refs. \onlinecite{Sellan2010}
and \onlinecite{Allen2014}, and clarified in Appendix \ref{App:PBD}. 
Second, an unexpected divergence (of the form $1/\sqrt{qa}$ begins to appear for cells
with small enough $q$ (relative to the transverse size $q_{\perp}\approx 2\pi/N_{y \ {\rm or} \ z}$.)  
Specifically, the onset of 
the upward turn appears roughly when $\sqrt(qa)<1/2N_y$, which corresponds to $N_x > 25N_y^2$,
a limit not always achieved in NEMD calculations.  The origin and significance of this
divergence is discussed in Appendix \ref{app:B}.  The idea for extrapolation is discussed in the
caption to Fig. \ref{fig:divergence15} and in Appendix \ref{app:C}.

\par
\begin{figure}
\includegraphics[angle=0,width=0.5\textwidth]{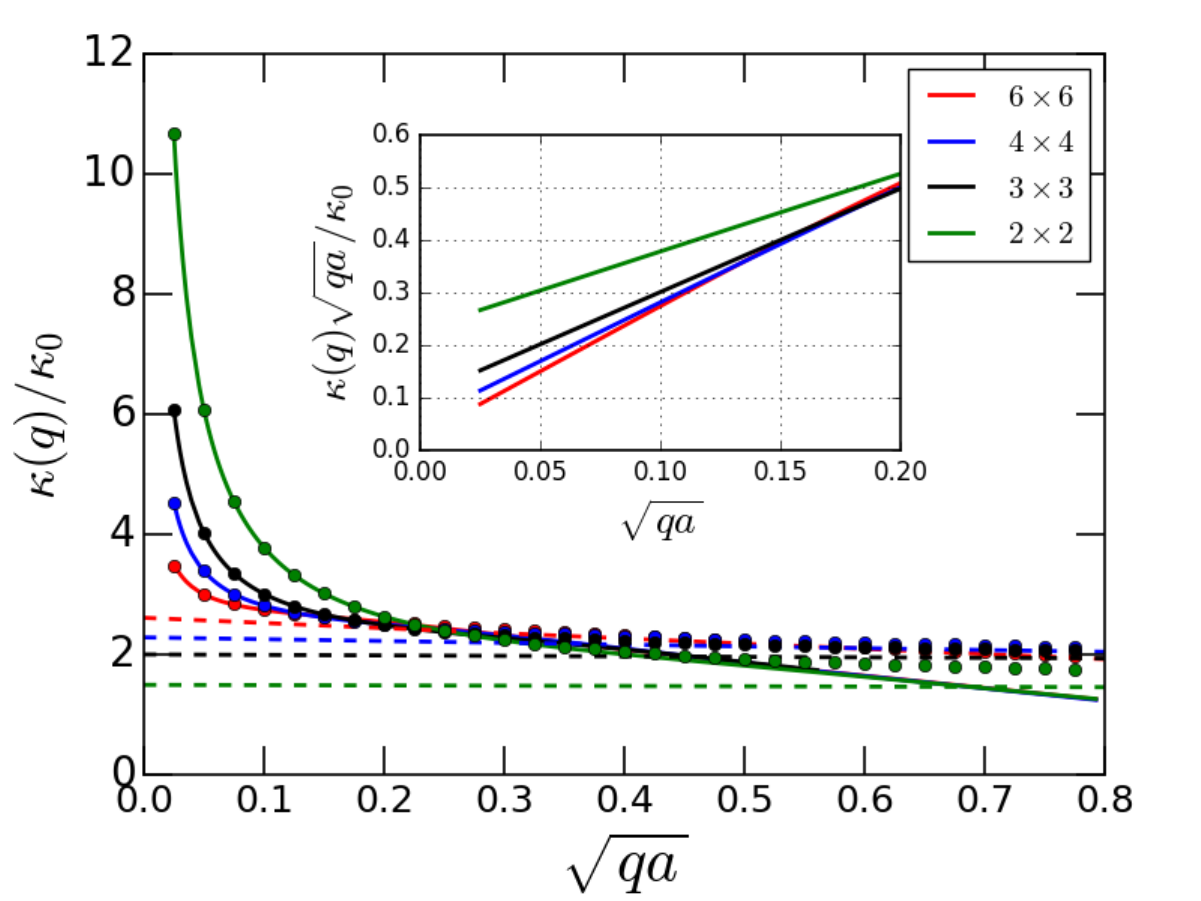}
\caption{\label{fig:divergence15} Convergence of $\kappa(q)$ toward $3\kappa_0$ found 
by numerical summation of the Boltzmann-Debye RTA model (with $p=2$). The simulation cell is anisotropic,
of size $N_y=N_z$ equalling 2, 3, 4, and 6 (top to bottom)  and $N_x=2\pi/qa$ varying from 10 to 10,000. 
The Q-sum in Eq.(\ref{eq:D1}) is evaluated not over the
Debye sphere, but over the anisotropic Q-mesh corresponding to the
normal modes of the {\it fcc} crystalline LJ lattice; results are shown as solid lines.
The inset shows that the small $q$ divergence has a $1/\sqrt{qa}$ form
with coefficient diminishing as $N_y$ increases.
The simulation results are fitted 
(for $50<N_x <1000$ or $.08<\sqrt(qa)<.35$) to a 3-term form $\kappa(q)\approx A/\sqrt(qa)+B+C\sqrt(qa)$.
The fits are indicated by dots whose colors agree with the lines.
The value of $B$ (for $N_x \ge 6$) is a reasonable choice for $\kappa(q=0)$.  The
dashed lines  show the non-diverging parts $\kappa(q) \approx B+C\sqrt(qa)$ of the 3-term fits.
The coefficient $B$ is the slope of the lines in the inset, and the $\sqrt(qa)=0$ intercept of the dashed lines
on the main graph.}
\label{fig:divergence15}
\end{figure}
\par


\section{Lennard-Jones crystal}
\label{sec:LJC}

Unlike the LJ liquid, for an LJ crystal, phonon gas theory
applies well, but only up to half the melting temperature, where
higher-order anharmonic terms become important \cite{Turney2009}.
This non-Boltzmann regime is where an MD simulation is worth doing .
The resulting shorter phonon mean free paths permit shorter simulation cells.
We simulate crystalline LJ argon at $T$=80K, close to the experimental
triple point (84K and 0.7 atmospheres) and atmospheric pressure melting temperature (84K).

Higher energy phonons have mean free paths a bit
longer than the unit cell $a = 5.32 \AA$, which we choose to be the slab thickness $d$. 
Lower energy phonons have mean free paths $\Lambda_Q$ increasing,
roughly as $1/\omega_Q^2$.  The values of $\Lambda_Q$ are not as long as in GaN, 
modeled by Zhou {\it et al.} \cite{Zhou2009}.  Nevertheless, 
doing a converged calculation by MD methods is challenging.  
We use this to test whether our algorithm is helpful.   
We use a time step of the ``velocity Verlet''  Newton's-law 
integration algorithm \cite{Hansen1969,Tuckerman1991} $\delta t = 0.007 t_{LJ}$ for smaller-size samples,
and $0.014 t_{LJ}$ for larger ones.

\par
\begin{figure}
\includegraphics[angle=0,width=0.5\textwidth]{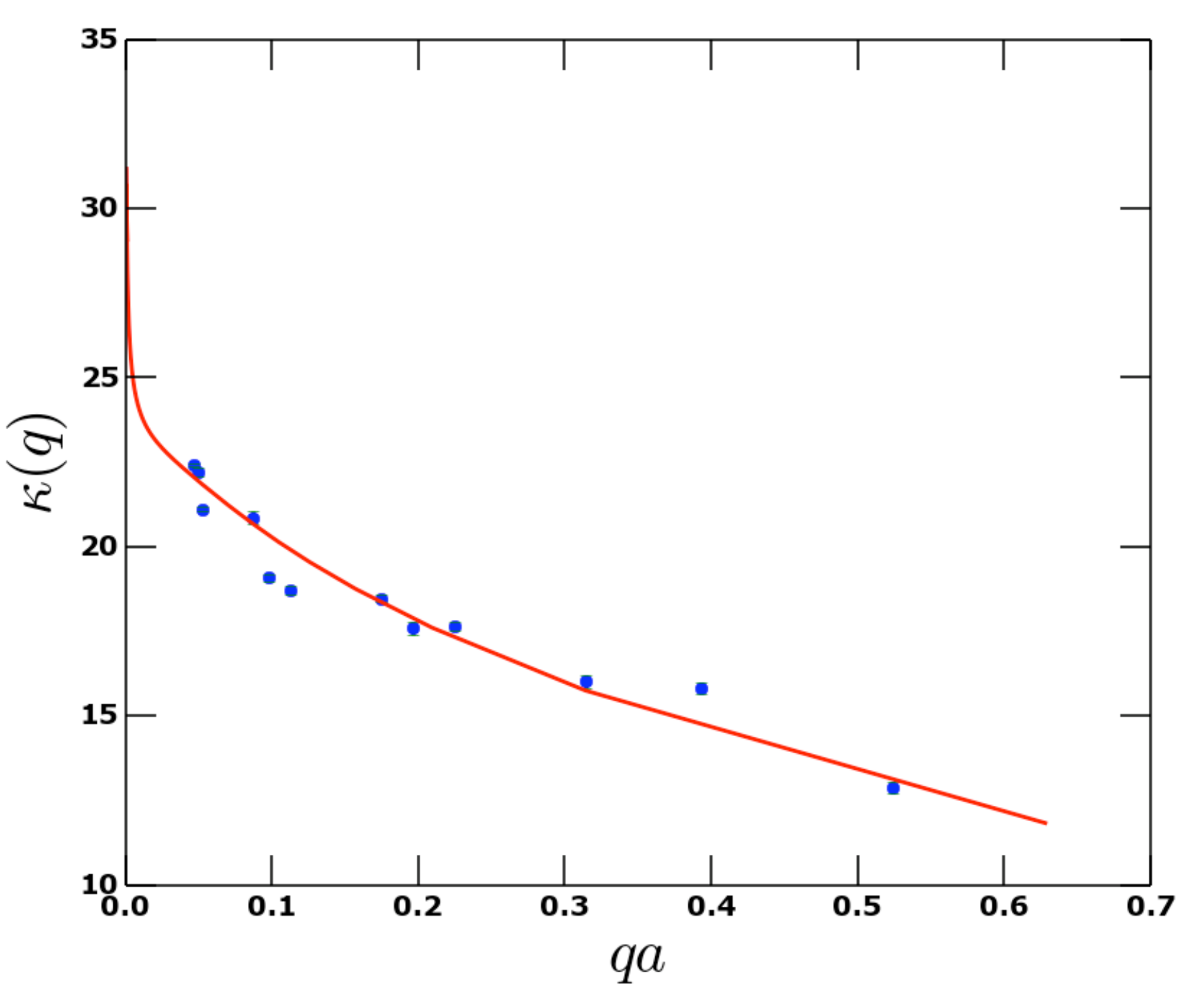}
\caption{\label{fig:powerlaw} Points are results of NEMD for the LJ crystal. 
Values of $\kappa$ are shown in LJ units ($\kappa_{\rm LJ}=(k_B/\sigma^2)\sqrt(\epsilon/M)$=
0.0188 W/mK). 
 The wave-vector $q=2\pi/L_x$ is the smallest compatible with the simulation cell,
whose size is $(L_x=N_x a) \times 6a \times 6a $.  Values of $N_x$ vary from 12 to 120. 
The red line is a numerical summation of the Debye-RTA model,
Eq. \ref{eq:D1}, with exponent $p=2$,
using a $Q$-mesh compatible with the
discrete normal mode quantum numbers of the $y,z$ size of the simulation cell, and $N_x=2\pi/qa$ 
values varying from 10 at the right, up to 10$^4$ at the left.  The
parameters $\kappa_0$ and $\tau_D$ are scaled to fit the NEMD numbers.  The
anomalous small-$q$ behavior is better seen in Fig. \ref{fig:error1}, and is shown in Appendix
\ref{app:B} to be an artifact of the coarse $N_{y,z}$ mesh.}
\label{fig:powerlaw}
\end{figure}
\par

\par
\begin{figure}
\includegraphics[angle=0,width=0.5\textwidth]{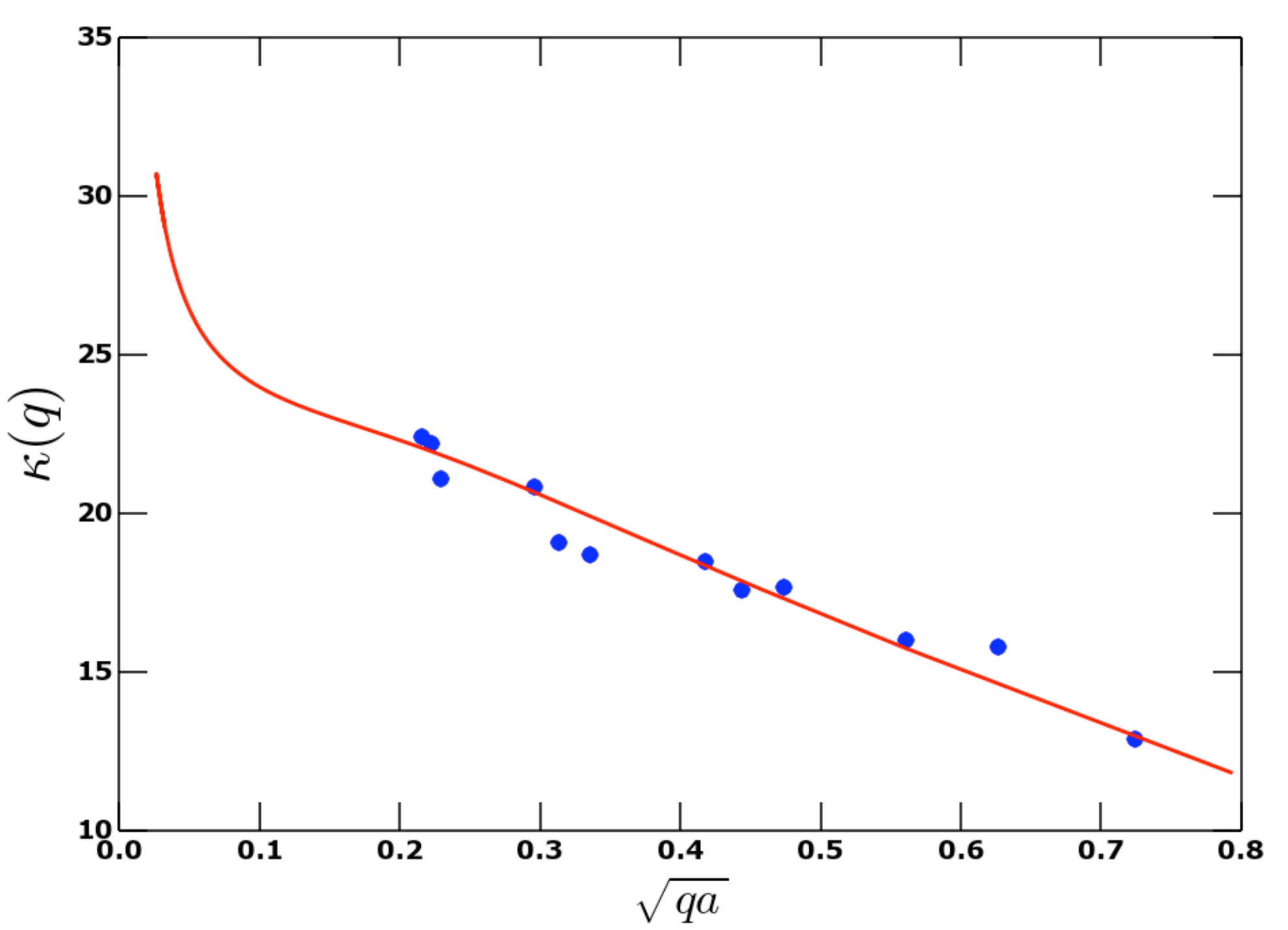}
\caption{\label{fig:error1} Points are numerical $\kappa(q)$ in LJ units,
the same as those in Fig. \ref{fig:powerlaw}.
 The red line is also as in Fig. \ref{fig:powerlaw},
a scaled RTA-Debye model with $p=2$ (Eq. \ref{eq:D1})
summed using the same coarse transverse $Q$-mesh as the simulation cell,
and varying $N_x$ and thus $q=q_{\rm min}=2\pi/L=2\pi/N_x a$.
Use of $\sqrt(qa)$ for the horizontal coordinate gives better linearity and a more reliable 
extrapolation to $q=0$.  The anomalous low-$q$ behavior of the red line is discussed
in Appendix \ref{app:B}.}
\label{fig:error1}
\end{figure}
\par

Figs.\ref{fig:powerlaw} and \ref{fig:error1} show results for $\kappa(q=2\pi/L)$ where $L=N_xd$ is the
length of the simulation cell, and $d=a$.  These calculations use a heat input $H(\ell)\Delta t$ (per slab)
of $1.265\epsilon\cos(2\pi\ell/N_S)$.  The interval $\Delta t$ between heat 
insertions is $60\delta t$.   
The value of $\kappa$ in argon at $T=$80 K is measured \cite{Clayton1973} to be in the
range 0.4 to 0.6 W/mK.  Christen and Pollack \cite{Christen1975} found $\kappa(80K)\approx 0.30$
W/mK.  Fig. \ref{fig:powerlaw} looks as if it might extrapolate linearly in the
region $qa<0.1$ to a value around 23 in LJ units, whereas Fig. \ref{fig:error1} seems
more convincingly linear in $\sqrt(qa)$, extrapolating to a value $\kappa(q=0)\approx$ 26 in
LJ units.  The LJ unit of thermal conductivity is $(k_B/\sigma^2)\sqrt(\epsilon/M)$=0.0188 /W/mK.
Our LJ crystal simulation thus gives $\kappa\approx$ 0.49 W/mK.  This compares with the
value $0.16-0.17$ W/mK found by Turney {\it et al.} \cite{Turney2009},  0.236 W/mK found by
Omini and Sparavigna \cite{Omini1995}, 0.25 W/mK found (at 69.2K) by Kaburaki {\it et al.} \cite{Kaburaki2007},
$0.33-0.49$ W/mK (depending on density) found by Christen and Pollack \cite{Christen1975}, and 
0.19 W/mK found by Chernatynskiy and Phillpot \cite{Chernatynskiy2010}. 


\section{conclusions}

The algorithm of Sec. \ref{sec:Sine} works  smoothly and converges more rapidly
than the common two-slab heating.   It generates a reliable value of $\tilde{\kappa}(q=2\pi/L)$.
The macroscopic conductivity, $\kappa={\rm Lim}_{q\rightarrow 0}\tilde{\kappa}(q)$, found by extrapolating
the long sample dimension to $L\rightarrow\infty$, is problematic, although less so than for
the discrete-slab heating algorithm.  Even if Boltzmann transport theory fails because
phonon mean free paths are so short that quasiparticles are not well-defined, 
nevertheless, Boltzmann theory should correctly model the long-wavelength phonon
contribution to $\kappa(q)$, which is the problematic part.

Our analysis using the BTE reveals two effects responsible for slower than expected convergence of
$\tilde{\kappa}(q)$ to $\kappa$.  These are an inevitable correction which scales as $\sqrt(qa)$,
and the anisotropic mesh artifact that scales as $1/[N_y N_z \sqrt(qa)]\propto \sqrt(N_x)/N_y N_z$.
These are found by numerical summation of the Boltzmann-Debye RTA version, but
should faithfully model the effects seen in NEMD models.  
Simple graphical extrapolation
to $q=0$ by assuming linear behavior in $\sqrt{qa}$ is less justifiable than had been hoped, because
the contamination by the $1/\sqrt(qa)$ term alters the appearance of the $\kappa(q)$ {\it versus}
$\sqrt(qa)$ graph.  The cure is to be sure that the ratio $\sqrt(N_x)/N_y N_z$ does not get too small.
Appendix \ref{app:C} discusses this further.  There are
reasons for mild insecurity on the issue of to what extent extrapolation is justified.


%
\begin{appendix}

\section{Peierls-Boltzmann-Debye (PBD) models}
\label{App:PBD}

The normal Debye model visualizes three acoustic branches of normal modes $Q$.
For simplicity, they are all given the same velocity $v$, and remain dispersionless
throughout the sphere of radius $Q_D$ which models the Brillouin zone.  For
anharmonic scattering, the relaxation rate $1/\tau_Q$ is $Q$-dependent.  An 
appropriate extension of the Debye model is to take $1/\tau_Q=1/\tau_D (Q/Q_D)^p$,
where the power $p$ is important but a bit uncertain in reality.  The result is
a family of Peierls-Boltzmann-Debye relaxation-time approximations (PBD-RTA). 
\begin{equation}
\tilde{\kappa}_p(q)=\frac{9}{2}\kappa_0 \int_0^1 dx x^2 \int_{-1}^1 d\mu \frac{\mu^2}{x^p + i\lambda\mu}
\label{eq:kp}
\end{equation}
where $\lambda=qv\tau_D=q\ell_{\rm min}$, and $Q$ is the phonon mode label.  The
integration is over $x=Q/Q_D$, and
$\mu=\cos\theta=Q_x/Q$.  The scale factor is $\kappa_0=k_B v^2 \tau_D/\Omega_{\rm cell}$.
This equation is just Eq.(13) of Ref. \onlinecite{Allen2014}.  It is also the continuum
version ($qd\rightarrow 0$) of the discrete slab Eq. \ref{eq:kappaslab1}.

For integer $p$, the integrations can be done analytically:
\begin{equation}
\frac{\tilde{\kappa}_0(q)}{\kappa_0}=\frac{3}{\lambda^2}\left[1-\frac{\tan^{-1}(\lambda)}{\lambda} \right]
\label{eq:k0}
\end{equation}
\begin{equation}
\frac{\tilde{\kappa}_1(q)}{\kappa_0}=\frac{9}{10}\left[ 1+\frac{2}{\lambda^2}
\left(1-\frac{\tan^{-1}(\lambda)}{\lambda}\right)
-\lambda^2 \log\left(1+\frac{1}{\lambda^2}\right)\right]
\label{eq:k1}
\end{equation}
\begin{eqnarray}
\frac{\tilde{\kappa}_2(q)}{\kappa_0}&=&\frac{9}{7\lambda^2}\left[ 1-\frac{\tan^{-1}(\lambda)}{\lambda}
+2\lambda^2 \right. \nonumber \\
- \lambda^2 \sqrt{\frac{\lambda}{2}} &&\left(\tan^{-1}\left(\sqrt{\frac{2}{\lambda}}+1\right)
+\tan^{-1}\left(\sqrt{\frac{2}{\lambda}}-1\right)\right) \nonumber \\
&-&\left. \frac{\lambda^2}{2} \sqrt{\frac{\lambda}{2}} \log\left(\frac{1+\sqrt{2\lambda}+\lambda}
{1-\sqrt{2\lambda}+\lambda}\right)\right]
\label{eq:k2}
\end{eqnarray}
\begin{equation}
\frac{\tilde{\kappa}_3(q)}{\kappa_0}=\frac{1}{2}\log\left(\frac{1+\lambda^2}{\lambda^2}\right)
+\frac{1}{\lambda^2}\left(1-\frac{\tan^{-1}(\lambda)}{\lambda}\right)
\label{eq:k3}
\end{equation}
These equations are plotted versus $\lambda=q\ell_{\rm min}$ in Fig. \ref{fig:kp}.  Of these formulas,
the most useful is probably the $p=2$ case, Eq.(\ref{eq:k2}).  It is also the most difficult to derive.
Details of the derivation are given in the Online Supplemental Material.

\par
\begin{figure}
\includegraphics[angle=0,width=0.5\textwidth]{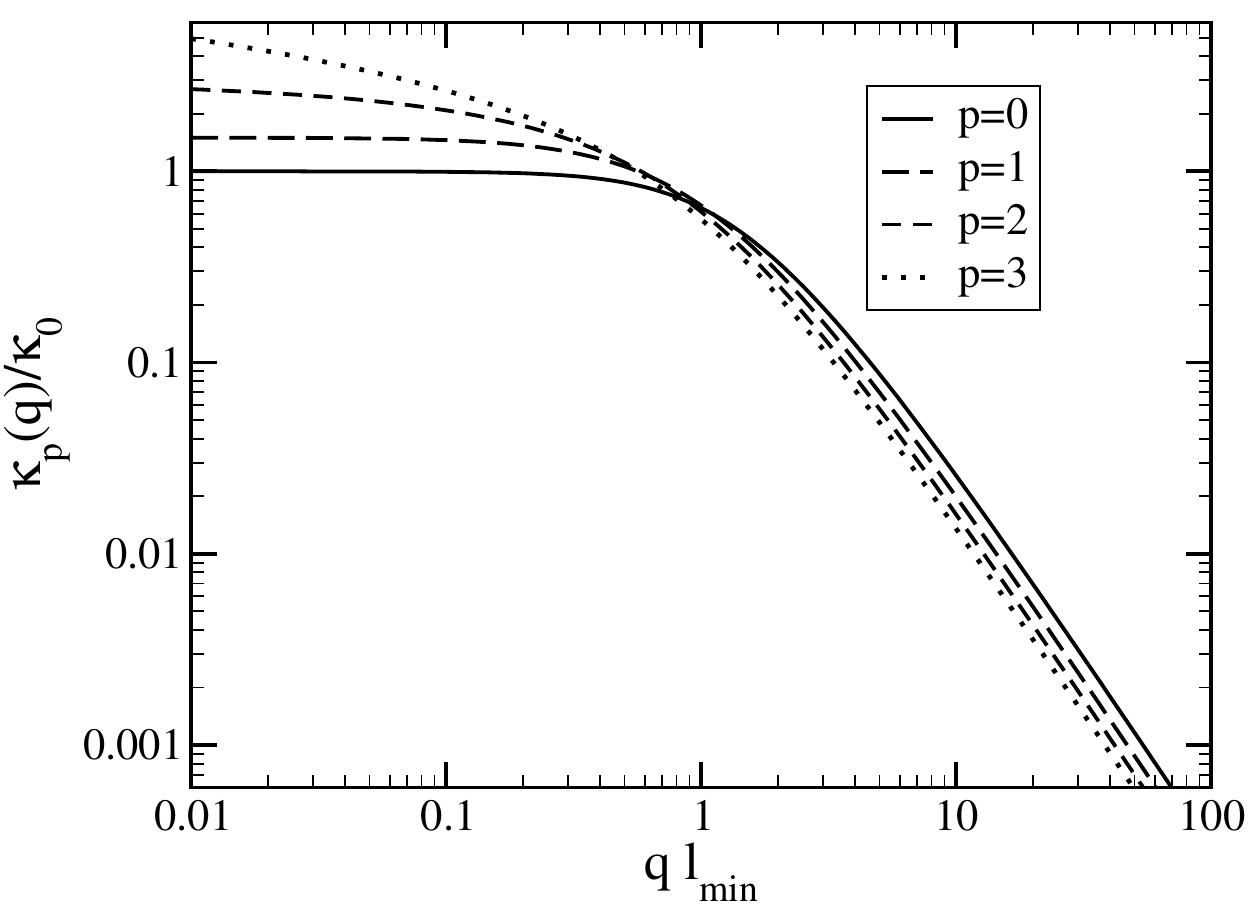}
\caption{\label{fig:kp} The $q$-dependent PBD-RTA conductivities $\tilde{\kappa}_p(q)$
as expressed in Eqs.\ref{eq:k0} to \ref{eq:k3}.
}
\label{fig:kp}
\end{figure}
\par

These four functions have simple large-$q$ limits, 
\begin{equation}
\tilde{\kappa}_p(q)=\frac{9}{3+p}\frac{\kappa_0}{(q\ell_{\rm min})^2}.
\label{eq:kpbigq}
\end{equation}
Their small-$q$ limits are diverse, and contain the rules for extrapolation to $q=0$.
The $p=0$ function $\tilde{\kappa}_0$ hardly requires extrapolation if the
cell size significantly exceeds $\ell_{\rm min}=v\tau_D$.  The $p=3$ function
diverges to infinity as $\log(1/q)$ when $q\rightarrow 0$.  If the exponent $p$
exceeds 3, the divergence is faster than logarithmic.  The explicit small $q$ expansions are
\begin{equation}
\frac{\tilde{\kappa}_0(q)}{\kappa_0}\rightarrow 1-\frac{3}{5}\lambda^2+\ldots
\label{eq:k00}
\end{equation}
\begin{equation}
\frac{\tilde{\kappa}_1(q)}{\kappa_0}\rightarrow \frac{3}{2} - \frac{9}{5}\lambda^2 \log(1/\lambda)
-\frac{9}{25}\lambda^2 +\ldots
\label{eq:k10}
\end{equation}
\begin{equation}
\frac{\tilde{\kappa}_2(q)}{\kappa_0}\rightarrow 3-\frac{9\pi}{7}\sqrt{\frac{q\ell_{\rm min}}{2}}+\ldots
\label{eq:k20}
\end{equation}
\begin{equation}
\frac{\tilde{\kappa}_3(q)}{\kappa_0}\rightarrow \log(1/q\ell_{\rm min})+3+\ldots
\label{eq:k30}
\end{equation}
For $p>0$, the leading term is non-analytic in $q$, but higher corrections are analytic.

\section{Discrete slab heating}
\label{App:keff}

The previous appendix \ref{App:PBD} deals with non-local $\tilde{\kappa}(q)$ for an infinite homogeneous crystal.
Then $q$ is a continuous variable with no upper bound.  Now we must deal with a situation
where an artificial superlattice periodicity is imposed, which forces $T(x+L)=T(x)$.  This means
that $q$ is no longer continuous, but quantized ($2\pi n/L$ for integer $n$).  Furthermore,
$T(x)$ is no longer defined for a continuous spatial variable $x$, but only at discrete values
$x(\ell)=\ell d$ where $d = L/N_S$.  The distance $d=n_S a$ is a multiple of the crystalline period 
$a$ in the $\hat{x}$-direction,
and $N_S=N_x/n_s$ is the total number of discrete slabs within which the temperature $T(\ell)$ is thermally averaged.
This both simplifies and complicates the Fourier analysis.  The simplification is that now there are
only a finite number ($N_S$) of Fourier coefficients for the functions of periodicity $L$.  Specifically,
$q=2\pi n/L = (2\pi/d)(n/N_S)$, where
$-N_S/2+1 \le n \le N_S/2$, or $-\pi/d < q \le \pi/d$.
The complication is that the finite domain of $q$ introduces simple but unfamiliar detailed 
differences from continuous cases.  Equation \ref{eq:nonlocal} is a good example.

Since discrete variables $T$ and $H$ are defined for distinct slabs $\ell$, the discrete variables $J$
and $dT/dx$ are defined between slabs ($\ell+1/2$, for example.)   For example, the Fourier representation of $J$ is
\begin{eqnarray}
&&\tilde{J}(q)=\frac{1}{N_S} \sum_\ell^{N_s \ {\rm values \ of} \ \ell} e^{-iqd(\ell+1/2)} J(\ell+1/2)  \nonumber \\
&&J(\ell+1/2) = \sum_q^{N_s \ {\rm values \ of} \ n} e^{iqd(\ell+1/2)} \tilde{J}(q).
\label{eq:Jq}
\end{eqnarray}
In the case where heat is added in the slab $\ell=0$ and removed in the slab $\ell= N_S/2$
(equivalent to slab $-N_S/2$), the current is $+J$ to the right and -J to the left of $\ell=0$, or
\begin{eqnarray} 
\tilde{J}(q)
&=& \frac{J}{N_S}\sum_{\ell=0}^{N_S/2-1} \left[e^{-iqx(\ell+1/2)} - {\rm c.c.} \right] \nonumber \\
&=& \frac{J}{N_S} \left( \frac{1-e^{-iqL/2}}{1-e^{iqd}} \right) e^{-iqd/2} - {\rm c.c.} \nonumber \\
&=& \frac{J}{iN_S} \left[ \frac{1-(-1)^{n_q}}{\sin(qd/2)} \right].
\label{eq:Jdph}
\end{eqnarray}
At $q=0$, the expression $[ \ ]$ in the last line 
of Eq.(\ref{eq:Jdph}) needs definition; the correct value is 0, as is also true for all $q$'s with even $n_q$.
This happens because of choices that made $J$ antisymmetric around the points of maximum
heat insertion ($\ell=0$) or removal ($N_S/2$).
Now let us analyze the approximate thermal conductivity,
\begin{equation}
\kappa_{\rm eff}(L) \equiv -J/\nabla_z T({\rm mid}).
\label{eq:keff}
\end{equation}
That is, the thermal conductivity is approximated by the ratio of the actual current $J$, controlled by the
heating rate $\dot{e}$, to the temperature gradient $-\nabla_z T({\rm mid})$ found midway between the 
heat input slab ($\ell=0$) and output slab ($\ell=N_S/2$).  This temperature gradient has the Fourier 
representation
\begin{eqnarray}
\nabla_z &T& ({\rm mid}) = \sum_q e^{iqN_Sd/4} \tilde{\nabla}_z T(q) \ \ [N_S/4={\rm half \ integer}] \nonumber \\
&=& \sum_q e^{iqN_Sd/4} \cos(qd/2) \tilde{\nabla}_z T(q) \ \ [N_S/4={\rm  integer}]. \nonumber \\
\label{eq:nabTdph}
\end{eqnarray}
In the case $N_S/4 = {\rm integer}$, the slab $\ell=N_S/4$ lies midway between heat input and output,
so the temperature gradient (needed at the slab mid-point) is taken as the average of the left
and right slab boundaries.  This introduces the factor $\cos(qd/2)$ in the second version of Eq.(\ref{eq:nabTdph}).
Finally, we substitute $\tilde{\nabla}_z T(q) = -\tilde{J}_z(q)/\tilde{\kappa}(q)$ in Eq.(\ref{eq:nabTdph}), and use 
Eq.(\ref{eq:Jdph}) for $\tilde{J}_z(q)$.  Then Eq.(\ref{eq:keff}) becomes
\begin{eqnarray}
\frac{1}{\kappa_{\rm eff}} &=& \frac{2}{N_S}\sum_q^{n_q={\rm odd}} \frac{\sin(qL/4)}{\sin(qd/2)} 
\left[ \frac{1 \ {\rm or} \ \cos(qd/2)}{\tilde{\kappa}(q)} \right] \nonumber \\
&=& \left[ \frac{4 \ {\rm or} \ 4\cos(\pi/N_S)}{N_S\sin(\pi/N_S)} \right] \frac{1}{\tilde{\kappa}(q_{\rm min})}
+\{|n_q| \ge 3 \ {\rm terms}\}. \nonumber \\
\label{eq:kqeff}
\end{eqnarray}
This is a surprisingly complicated relation between the size-dependent ``computational'' value
of $\kappa_{\rm eff}$ and the Fourier representation $\tilde{\kappa}(q)$.   
From Eq.(\ref{eq:kqeff}), the leading term (at small $q_{\rm min}d/2=\pi d/L=\pi/N_S$)
is $1/\kappa_{\rm eff}=4/\pi\tilde{\kappa}(q_{\rm min})$, with oscillatory
corrections $~\sum_{n=1} (4/\pi) (-1)^n /[(2n+1)\tilde{\kappa}((2n+1)q_{\rm min})]$.  
In the local limit, $\tilde{\kappa}(q)=\kappa$,
Eq.(\ref{eq:kqeff}) converges exactly to $\kappa_{\rm eff}=\kappa$.


\section{Anisotropic mesh}
\label{app:B}

Fig. \ref{fig:error1} indicates that Eq.(\ref{eq:k20}) gives a good match
to the size-dependence when $N_x$ is not too big (and $q$ is not too small.)  At smaller
$q$ there is an up-turn in the numerical Debye-RTA sum,
that is not derived in the analytic integration Eq. \ref{eq:k2}.
This up-turn is strongly enhanced at small transverse cell size $N_yN_z$.
The problem is that the ratio
$L_x/L_z$ becomes very large at small $q=2\pi/L_x$.
It is necessary to reconsider how Eq. \ref{eq:D1} (for $p=2$) behaves in a finite-size crystal or simulation cell,
when one dimension ($L_x\equiv L$) gets large but the other two ($L_y = L_z \equiv L_{yz}$)
remain small.  The answer is, a new non-analytic piece occurs.  The results in Appendix \ref{App:PBD}
are for a crystal with size going to $\infty$ in all directions.   For $p=2$, $\tilde{\kappa}(q)$ deviates from  
$\kappa$ as $\sqrt q$ at small $q$.  Here we show that the deviation becomes like $1/\sqrt{q}$
if the simulation cell has too large a ratio of $N_x$ to $N_y N_z$.
This is specific to the power law relaxation 
$\tau_Q=\tau_D(\omega_D/\omega)^p$ with $p=2$.  The divergence is a property of a one-dimensional
wire, indicating that ballistic transport has a dominant effect in such a system.
As the area $L_y L_z$ of the wire increases, the divergent term in $\kappa(q_{\rm min})$
decreases as $a^2/L_y L_z$, restoring the $\sqrt q$ answer.

The specific system under consideration is an fcc crystal of volume $Na^3 /4$, where
$a$ is the conventional primitive cube size, $N=4N_x N_y N_z$ is the total number of
atoms, $N_y$ and $N_z$ being small integers held fixed, and $N_x$ being a large
integer.  We seek the behavior as $N_x$ gets very large.
The conductivity is given by Eq. \ref{eq:D1}, rewritten as 
\begin{equation}
\frac{\tilde{\kappa}(q_{\rm min})}{\kappa_{0}}=\frac{1}{N} \sum_{Q_x} \sum_{Q_y,Q_z} 
 \frac{ (Q_x/Q)^2 (Q_D/Q)^2 } {  1 +(Q_x/Q)^2  \lambda^2 (Q_D/Q)^4  },
\label{eq:B1}
\end{equation}
where $\lambda=qv\tau_D$. 
The $Q$-vectors are $\vec{Q}=(2\pi/a)(n_x/N_x,n_y/N_y,n_z/N_z)$.  This choice is required
to make vibrational normal modes satisfy periodicity in the supercell.  There are
$N$ $Q$-vectors in the Brillouin zone.
The Debye model simplifies summation over the Brillouin zone by using the Debye
sphere, with a volume equal to $n$ times the volume of the primitive Brillouin zone, $n$
being the number of atoms in the primitive cell, 4 for fcc.  The $Q$-points are
dense along the $x$ direction and sparse in the others.  Only the $Q_x$-sum
can be turned into an integral.  It is consistent with the philosophy of the Debye
model, to not use a sphere in this case, but a cube-shaped ``pseudo-Brillouin zone'', of 
volume $4$ times $(2\pi/a)^3$, containing the correct number of states.  
The $Q_x$ sum is then an integral, going from 
$Q_x=0$ to the boundary of the ``pseudo-Brillouin zone,'' $4^{1/3}\pi/a$, and multiplied by 2
to cover both negative and positive $Q_x$.  
\begin{eqnarray}
\frac{\tilde{\kappa}(q_{\rm min})}{\kappa_{0}}&=&\frac{1}{4N_y N_z} 
\sum_{Q_y,Q_z} \frac{a}{\pi} \nonumber \\
&&\int_0^{4^{1/3}\pi/a} dQ
 \frac{ (Q_x/Q)^2 (Q_D/Q)^2 } {  1 +(Q_x/Q)^2  \lambda^2 (Q_D/Q)^4  }
\label{eq:B2}
\end{eqnarray}

The number of terms in the $Q_y,Q_z$ sum is $\approx n^{2/3}N_y N_z$, typically 50
for an MD simulation, or a few thousand for a small nanowire.  Of these terms, the one
which requires special attention is the $Q_y = Q_z = 0$ term.  We denote this
term by $\tilde{\kappa}_{00}(q_{\rm min})$,
\begin{eqnarray}
\frac{\tilde{\kappa}_{00}(q_{\rm min})}{\kappa_{0}}&=&\frac{1}{4N_y N_z} 
\frac{a}{\pi}\int_0^{4^{1/3}\pi/a} dQ \frac{Q^2 Q_D^2 } {  Q^4 + \lambda^2 Q_D^4  }\nonumber \\
&=& \frac{(3/\pi)^{1/3}}{2N_y N_z}\int_0^\zeta du \frac{u^2}{u^4 + \lambda^2},
\label{eq:B3}
\end{eqnarray}
where the variable $u$ is $Q/Q_D$, 
and the upper limit is $\zeta=(\pi/6)^{1/3}$.
The answer is
\begin{equation}
\frac{\tilde{\kappa}_{00}(q_{\rm min})}{\kappa_{0}}=\frac{(3/\pi)^{1/3}}{2N_yN_z}[g(\zeta)-g(0)],
\label{eq:B3a}
\end{equation}
where the indefinite integral $g(u)$ is
\begin{eqnarray}
g(u)&=&\frac{1}{2\sqrt(2\lambda)}
\left[ -\frac{1}{2}\log\left(\frac{u^2+\sqrt{2\lambda}u+\lambda}{u^2+\sqrt{2\lambda}u+\lambda}\right)\right.
\nonumber \\
&&+\left.\tan^{-1}\left(\frac{\sqrt{2\lambda}u}{\lambda-u^2}\right)\right].
\label{eq:B3b}
\end{eqnarray}
Because $N_x$ is large, $\lambda=2\pi v\tau/N_x a$ is small, and to first order, the definite
integral $g(\zeta)-g(0)$ is determined by $\tan^{-1}(\sqrt(2/ \lambda))\sim\pi/2$, and equals
$\pi/4\sqrt(2\lambda)\propto 1/\sqrt q$, insensitive to the details of the ``pseudo-Brillouin zone''
boundary location.  Thus to leading order, the piece of $\tilde{\kappa}(q)$ coming from the
$Q_y=Q_z=0$ part of the $Q$-mesh, is
\begin{equation}
\frac{\tilde{\kappa}_{00}(q_{\rm min})}{\kappa_{0}}= \frac{(3/\pi)^{1/3}}{2N_y N_z}
\frac{\pi}{8}\sqrt{\frac{L}{\pi v\tau}}
\label{eq:B4}
\end{equation}

The Debye model is reliable as a guide for the low frequency behavior, provided the
relaxation-time approximation and the associated power law $p$ are correct.  The
conclusion is a bit surprising.  It indicates that if anisotropic simulation cells are used
for NEMD simulation of $\kappa$, then the extrapolation to very long cells suffers
from an unintended 1D singularity.  The product $N_y N_z$ should increase at least
as rapidly as $N_x^{1/2}$ to prevent this term distorting the extrapolated answer.
In actual simulations, this is probably more a sobering thought than a serious warning.
But the effect is real, and shows up in the Debye-model numerics shown in Figs. 
\ref{fig:powerlaw} and \ref{fig:error1}.


%
\section{Semi-empirical Fitting}
\label{app:C}

\par
\begin{figure}
\includegraphics[angle=0,width=0.5\textwidth]{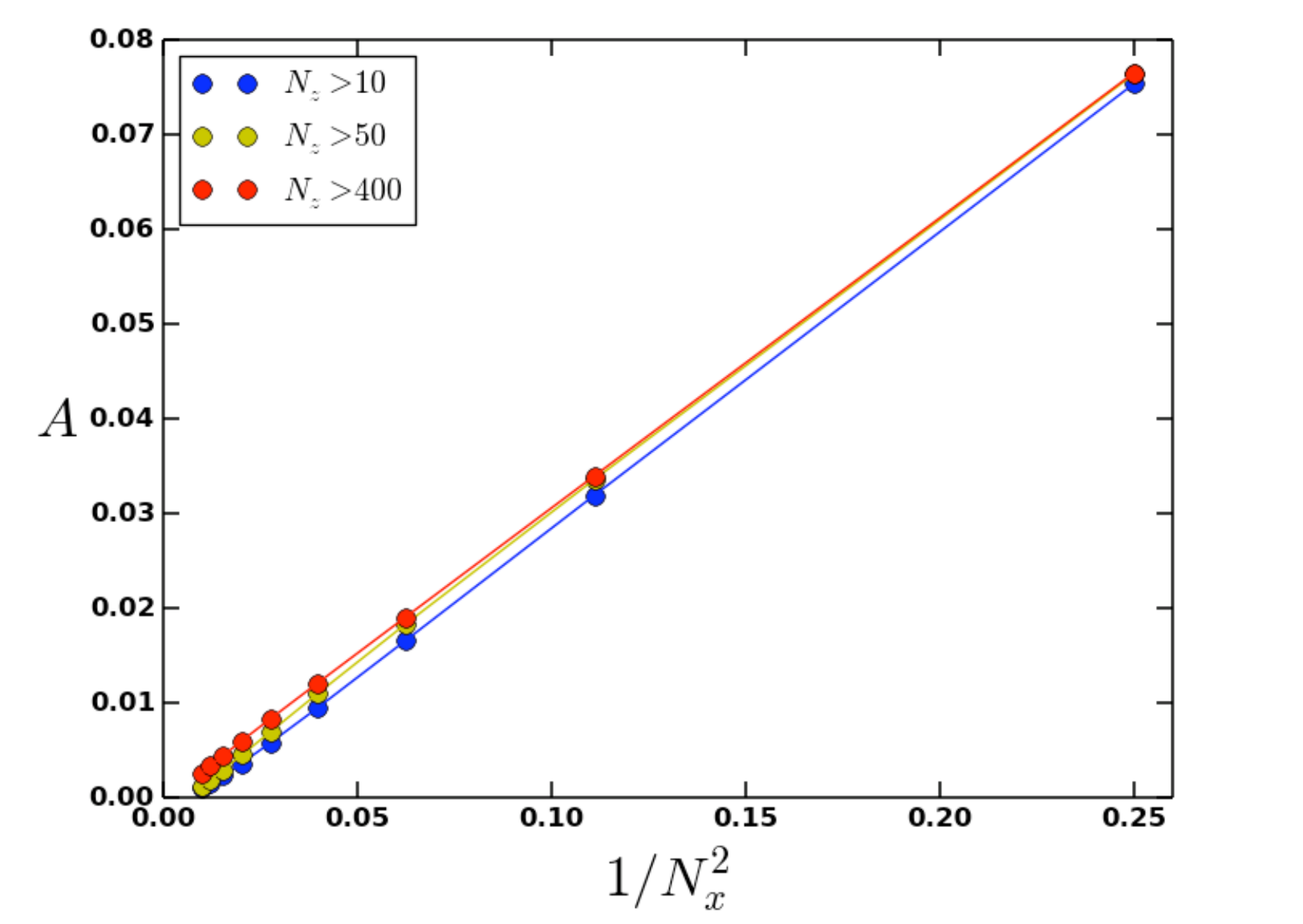}
\caption{\label{fig:A} The coefficient of the singular part $\propto 1/\sqrt(qa)$ scales nicely with 
the reciprocal of the transverse dimension $1/N_y^2=1/N_z^2$ of the simulation cell, especially
when the semi-empirical fit is restricted to smaller values of $q$.}
\label{fig:A}
\end{figure}
\par
\par
\begin{figure}
\includegraphics[angle=0,width=0.5\textwidth]{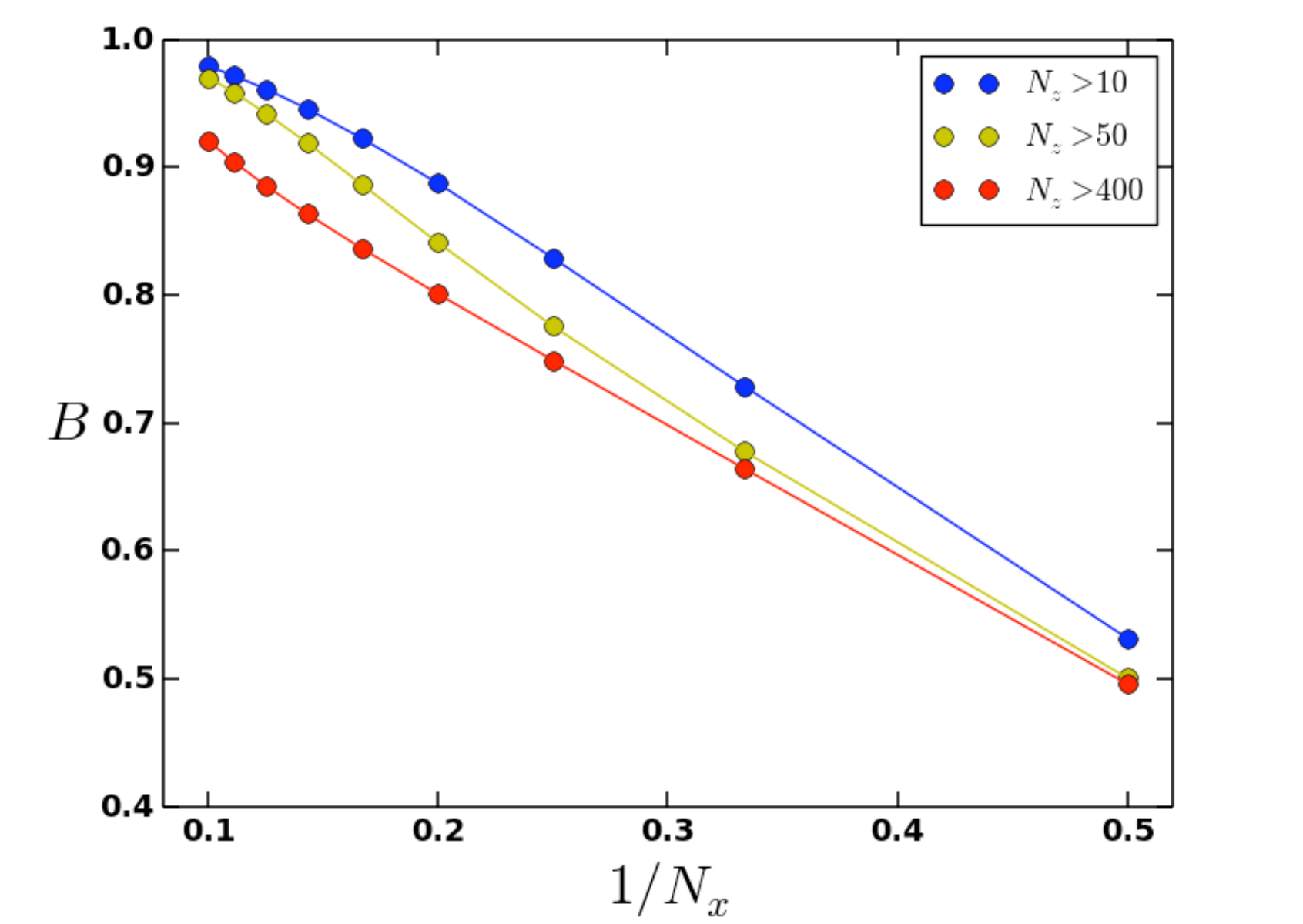}
\caption{\label{fig:B} The expected $q\rightarrow 0$ limit of Debye-RTA numerical sums corresponds to
coefficient $B\rightarrow 1$, which happens nicely as the transverse cell size is increased to $10 \times 10$.
The fact that it happens most quickly when even short cells $N_x = 10$ are included in the fit, suggests
fortuitous error cancellation.}
\label{fig:B}
\end{figure}
\par

Figures \ref{fig:powerlaw} and \ref{fig:error1} show (as red lines)  Debye-RTA discrete Q-sums 
adjusted to fit NEMD results.  These are intended to guide extrapolation, but reveal possible
ambiguity about how to correct for the $1/\sqrt(qa)$ behavior.  This is an artifact of too large
a ratio $N_x/N_y$, not achieved in the NEMD simulations, but easily achieved in the
Debye-RTA numerical sums on discrete $Q$-meshes.  
Here we show some results of a 3-term fit to the numerical Debye-RTA
sums.  From Appendices \ref{App:PBD} and \ref{app:B}, we are led to expect
behavior of the type
\begin{equation}
\kappa(q)/3\kappa_0 \approx A/\sqrt{qa} + B + C\sqrt{qa}.
\label{eq:C1}
\end{equation}
The coefficient $A$ should diminish as $N_y N_z$ increases, since it arises from only the $Q_y=0$ and $Q_z=0$
part of the normal mode spectrum, one part out of the total of $N_y N_z$  $Q_y$, $Q_z$-values.  This is tested
in Fig. \ref{fig:A}, and found correct.  Numerical summation over the $Q$-mesh was done for 7 choices 
of $N_y=N_z$, namely 2, 3, 4, 5, 6, 8, and 10, and for a mesh of $N_x$ ranging from the coarse value of 10 to 
the dense value of 10,000.
The 7 resulting $\kappa_{\rm Debye}(q)$ curves were least-squares fitted to Eq. \ref{eq:C1}.  The fitting was
done for the smallest $q=2\pi/N_x a$, up to a cutoff (all $N_x$ greater than a minimum value, chosen as 10, 50, or 200.)
The fit is extremely accurate for the smallest $q$-cut ($N_x=200$) and least accurate for the largest ($N_x=10$.)
The scaling with $1/N_y^2$ behaves just as expected.  The singular up-turn at very small $q$ is indeed an
artifact of an anisotropic simulation cell, and should be avoided by not letting the ratio
$N_x/(N_y N_z)$ become too large.

The coefficient $B$ is supposed to represent the true converged $q\rightarrow 0$ limit of 
$\kappa(q)$.  Figure \ref{fig:B} shows how this coefficient behaves as the transverse dimension increases.
For $N_y=N_z=10$, the Debye-RTA calculation is converged to 96\% or better, no matter what range
of $q$ is used for least-squares fitting.  Even down to $N_y=N_z=5$, the value of $B$ is 95\% of
the bulk value 1, provided all cell sizes down to the smallest ($N_x=10$) are included in the least-squares
fit, whereas, if only small $q$'s  ($N_x > 200$) are used, convergence is a lot slower.  This odd result
indicates that the improved convergence found by fitting the less relevant large $q$'s 
is likely to be partially a result of a fortuitous cancellation of errors.

\par
\begin{figure}
\includegraphics[angle=0,width=0.5\textwidth]{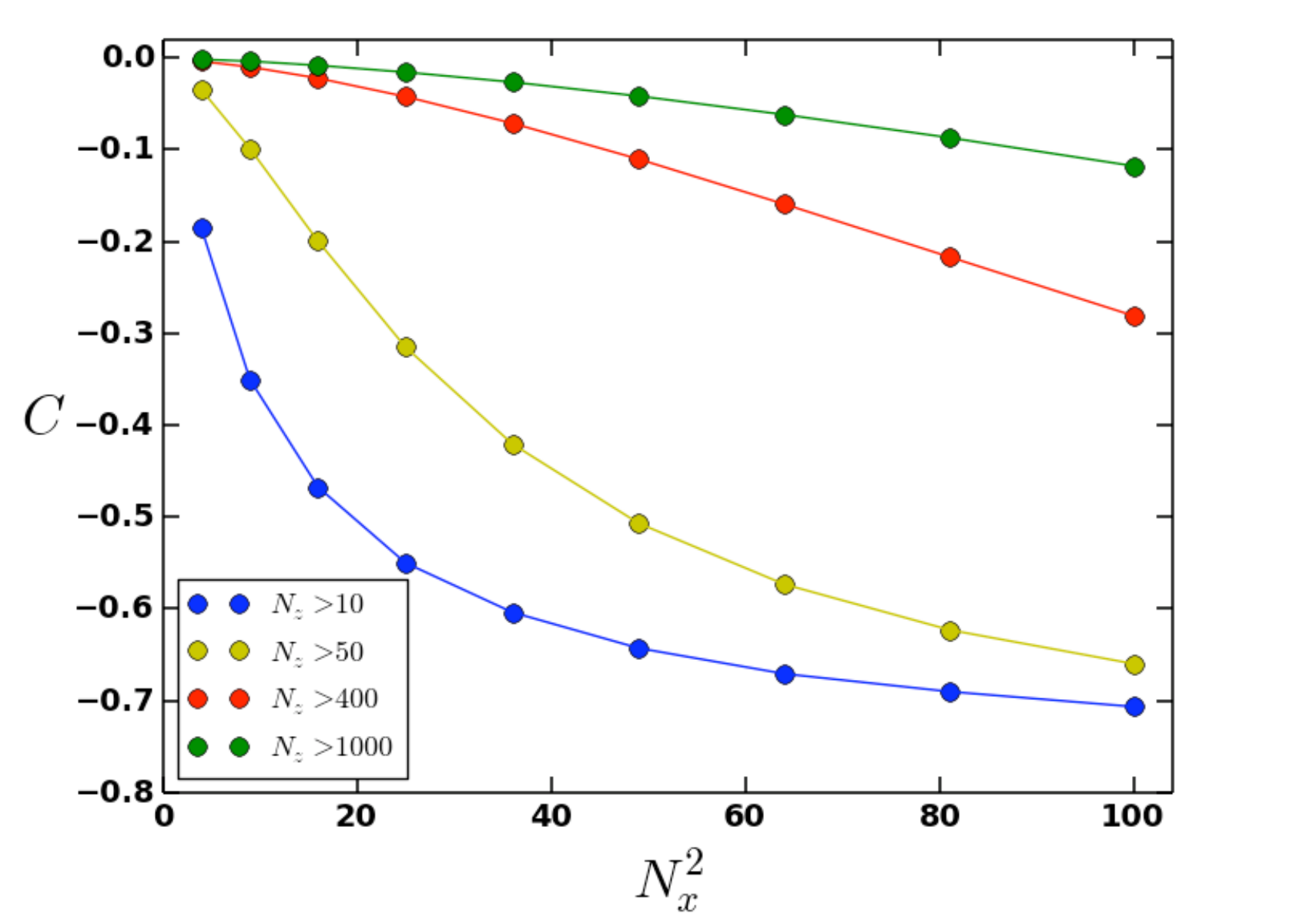}
\caption{\label{fig:C} The coefficient $C$ of the leading finite size correction $\kappa(q)/\kappa_0 \approx
1 + C\sqrt(qa)$ is suprisingly sensitive to transverse simulation cell size and to the choice of
$q$-range to use in the fit.}
\label{fig:C}
\end{figure}
\par

Finally, the leading finite size effect for bulk $\kappa(q)$ is contained in the term $C\sqrt(qa)$.
This term comes from long-wavelength phonons whose mean-free path exceeds the cell
size unless the cell is large in all three directions.  This negative contribution to $\kappa$ is apparently
suppressed when the cell becomes more anisotropic.  Simultaneously
the diverging term is getting larger.  These seem to accidentally compensate, making the NEMD
answers better than the true convergence estimates lead one to expect.  This is nicely illustrated in
Fig. \ref{fig:divergence15}, which shows that for $N_x$ not too big ($0.22<\sqrt(qa)<0.5$, corresponding
to $25<N_x<130$), the curve of $\kappa(q)$ {\it versus} $\sqrt(qa)$ is remarkably independent of $N_x$
and appears to point smoothly to the correct extrapolated value of $\kappa_0$.  The more careful fitting
to the 3-term expression gives extrapolated values shown by the dotted lines, which converge well for
meshes $N_y \ge 6$, but not as rapidly as the less careful extrapolation.  The prescription for extrapolation
of NEMD with realistic simulation cells seems to be, don't use meshes with $N_y = N_z$ much smaller than 6,
and extrapolate linearly if numerical results appear linear when plotted against $\sqrt(qa)\propto 1/\sqrt(N_x)$.

\end{appendix}
%


\section{acknowledgements}
We thank A. J. H. McGaughey for helpful advice.  We thank M. V. Fernandez-Serra
and J. Siebert for inspiration.
We thank the Stony Brook University Institute for Advanced Computational Science (IACS) 
for time on their computer cluster.
This work was supported in part by DOE grant No. DE-FG02-08ER46550.


\bibliographystyle{apsrev}
\bibliography{algori}

\begin{thebibliography}{46}
\bibitem[{foo({\natexlab{a}})}]{footnote1}
\bibinfo{note}{Recent literature often claims that thermal conduction in
  inhomogeneous systems disobeys the Fourier law. It would be better to say
  that it disobeys the local Fourier law
  $\kappa(\vec{r}-\vec{r}^\prime)=\kappa\delta(\vec{r}-\vec{r}^\prime)$, or
  $\tilde{\kappa}(\vec{q})=\kappa$. The same J. Fourier who realized
  $\vec{J}=-\kappa\vec{\nabla}T$ also invented the Fourier transforms used
  here.}
\end{thebibliography}

\end{document}


\title{On-line Supplemental Material for \\
Phonon thermal conductivity by non-local non-equilibrium molecular dynamics}

\author{ Philip B. Allen }
\email{philip.allen@stonybrook.edu}
\affiliation{ Department of Physics and Astronomy,
              Stony Brook University, 
              Stony Brook, New York 11794-3800, USA }

\author{ Yerong Li }
\email{yerong.li@outlook.com}
\affiliation{ Department of Physics and Astronomy,
              Stony Brook University, 
              Stony Brook, New York 11794-3800, USA }
\affiliation {Department of Intensive Instruction, Nanjing University, Nanjing 210093, China}

\date{\today}

\begin{abstract}
 
This supplement contains details of the FAT algorithm used to find the best value
of $\pm\Delta \vec{v}$ to use in the heating algorithm.  It also contains details of the
analytic integrations shown in Appendix A.

\end{abstract}

\maketitle


\section{Heat exchange algorithm}
\label{App:Furt}

The M\"uller-Plathe recipe is:
find the hottest atom in the cold ($\ell=N_S/2$) slab, and
the coldest atom in the hot ($\ell=0$) slab.  Interchange their velocities.  Energy and momentum
are both conserved, and the system is driven from equilibrium in a 
way that must be monitored.  Cao and Li \cite{Cao2010} among others,
have suggested modified algorithms of this type.  Furtado, Abreu, and Tavares\cite{Furtado2015} (FAT)
devised a gentler and more versatile method.  An earlier version of the present paper, which was
was posted on the arXiv\cite{Yerong2016}, derived the same procedure.  We call it the FAT algorithm,
because Furtado {\it et al.} discovered it simultaneously and published it first.

In our computation, driving is  done by spatially periodic injection ($\dot{e}(\ell)\propto\cos(2\pi\ell/N_s)$)
and simultaneous removal of heat.  Examine a 
slab $\ell$ (with $0 \le \ell < N_S/4$), and its conjugate slab $N_S/2-\ell$.  The former
is ``hotter'' than the latter because ${\dot e}(\ell)>0>{\dot e}(N_S/2-\ell)=-{\dot e}(\ell)$.  Find the 
coldest atom (meaning least kinetic energy) of all atoms of mass $m_i$ 
in the hotter slab, denoting its velocity as $\vec{v}_C$.  Find the hottest atom
of the same mass in the colder slab, denoting its velocity $\vec{v}_H$. 
Given the large fluctuations of the Maxwell-Boltzmann ensemble, it is certain that 
$v_H^2 > v_C^2$.  Choose an appropriate velocity $\vec{w}$ and
add it to the velocity $\vec{v}_C$ and subtract it from $\vec{v}_H$:
\begin{eqnarray}
\vec{v}_H &\rightarrow& \vec{v}_H - \vec{w} \nonumber \\
\vec{v}_C &\rightarrow& \vec{v}_C + \vec{w} 
\label{eq:changev}
\end{eqnarray}
The same operation should be done for the pair of slabs $-\ell$ and $\ell-N_S/2$.  All slabs can
be done simultaneously, or different random times can be used for different slabs.

There are three criteria for an appropriate $\vec{w}$, which uniquely fix the desired choice.
(i) The cold atom's kinetic energy should increase by $\Delta$, an energy that can be specified
in advance as $({\dot e}\Omega_S\tau)\cos(2 \pi \ell/N_S)$, where $\tau$ is the average time interval
between random interventions, and $\Omega_S$ is the volume of a slab.
(ii) The hot atom's kinetic energy should decrease by $\Delta$.  Then both momentum and 
energy are conserved.  Heating has the desired sinusoidal form, 
with ${\dot e}$ chosen not too different from $k_B T/\Omega_S\tau$.  Trial calculations
should test for the best choices of ${\dot e}$ and $\tau$.
(iii) There is still a one-dimensional family of vectors $\vec{w}$; from these, choose the smallest
$|\vec{w}|$, which gives the least impulse to the affected atoms.
Except for the sinusoidal spatial variation of $\Delta$, these are exactly the criteria 
chosen by Furtado {\it et al.}.  Their implementation of these criteria is also exactly like
ours.  We present our derivation here for the convenience of the reader.

\par
\begin{figure}
\includegraphics[angle=270,width=0.5\textwidth]{spheres.pdf}
\caption{\label{fig:spheres} Geometric construction for finding the smoothest velocity change $\vec{w}$
of the coldest atoms (with velocity $\vec{v}_C$  in the hotter region and the hottest atoms
(with velocity $\vec{v}_H$ in the colder region.  Both figures
represent a sphere of radius $\vec{r}=(\vec{v}_C -\vec{v}_H)/2$.  The plane perpendicular to {$\vec{b}$},
intersects the sphere in a circle, which is the locus of solutions $\vec{w}$ obeying energy and momentum
conservation rules.  The points $C$ and $C^\prime$ (in the $\vec{r}$--$\vec{b}$ plane) give the
solutions with least and greatest impulse.  The right-hand version shows the largest vector $\vec{b}$
which allows solutions for $\vec{w}$.  The minimum and maximum impulse solutions have
merged to a point.}
\end{figure}
\par

The energy shift criteria (i) and (ii) give equations 
$-2\vec{v}_H \cdot\vec{w} +w^2=-2\Delta/m\equiv-\delta$,
and $2\vec{v}_C \cdot\vec{w} +w^2=+\delta$.  Adding and subtracting these equations
give
\begin{eqnarray}
\delta&=&(\vec{v}_H + \vec{v}_C)\cdot \vec{w} \nonumber  \\
w^2 &=& (\vec{v}_H-\vec{v}_C)\cdot \vec{w},
\label{eq:geom}
\end{eqnarray}
a linear and a quadratic equation for $\vec{w}$.  These have a simple geometric interpretation
shown in Fig. \ref{fig:spheres}.  The first equation restricts the projection of $\vec{w}$ along 
the vector $\vec{v}_H + \vec{v}_C$.  Geometrically, this means that $\vec{w}$ lies on the plane
(shown by $C,C^\prime$ in Fig. \ref{fig:spheres})
perpendicular to the vector $\vec{b}\equiv \delta (\vec{v}_C + \vec{v}_H)/|\vec{v}_C+\vec{v}_H|^2$,
where the origin of $\vec{w}$ coincides with the origin of $\vec{b}$.
The second equation restricts $\vec{w}$ to the surface of a sphere of radius $|\vec{v}_H-\vec{v}_C|/2$,
centered at the end of the vector $\vec{r}\equiv(\vec{v}_H-\vec{v}_C)/2$, whose origin also coincides
with the origin of $\vec{w}$.  The sphere and the plane intersect on a circle, indicated by $C,C^\prime$
in Fig. \ref{fig:spheres}.   This circle is the one-dimensional family of solutions $\vec{w}$
satisfying Eqs.(\ref{eq:geom}).  It is also clear from the geometry that the shortest vector
$\vec{w}$ (the one that satisfies criterion (iii))
 is the one shown, from $O$ to $C$.  This lies in the same plane as the two known
vectors $\vec{r}$ and $\vec{b}$ (also the same plane as $\vec{v}_H$ and $\vec{v}_C$).  Therefore 
\begin{eqnarray}
\vec{w}&=&\alpha\vec{r} + \beta\vec{b} \nonumber \\ 
\vec{r}&=& (\vec{v}_H-\vec{v}_C)/2  \nonumber \\
\vec{b} &=& \frac{\delta(\vec{v}_H + \vec{v}_C)}{|\vec{v}_H+ \vec{v}_C|^2}.
\label{eq:defs}
\end{eqnarray}
These definitions allow Eqs.(\ref{eq:geom}) to be written as
\begin{eqnarray}
b^2&=&\vec{w}\cdot\vec{b} \nonumber  \\
w^2 &=& 2\vec{r}\cdot \vec{w}.
\label{eq:geom1}
\end{eqnarray}
The solution for $\vec{w}$ is
\begin{eqnarray}
\beta &=& 1-\alpha \frac{\vec{r}\cdot\vec{b}}{b^2} \nonumber \\
\alpha &=& 1-\sqrt{1-X}  \nonumber \\
X &=& \frac{(b^2-2\vec{b}\cdot\vec{r})b^2}{b^2 r^2 - (\vec{b}\cdot\vec{r})^2}
\label{eq:soln}
\end{eqnarray}
To derive this, substitute Eq.(\ref{eq:defs}) for $\vec{w}$ in terms of the
unknown coefficients $\alpha$ and $\beta$ into the Eqs.(\ref{eq:geom}).  The linear equation
is used to find $\beta$ in terms of $\alpha$.  Eliminating 
$\beta$ in favor of $\alpha$ in the quadratic equation gives a quadratic equation for
$\alpha$.  The appropriate solution is displayed in Eq.(\ref{eq:soln}).
An alternate version directly in terms of the velocities $\vec{v}_H$ and $\vec{v}_C$ is
\begin{equation}
X= (2\Delta/m) \frac{(2\Delta/m)-(v_H^2-v_C^2)}{v_H^2 v_C^2- (\vec{v}_H\cdot\vec{v}_C)^2}
\label{eq:X}
\end{equation}
\begin{equation}
\vec{w}=\frac{\alpha}{2}(\vec{v}_H-\vec{v}_C) + \left[\frac{2\Delta}{m}-\frac{\alpha}{2}(v_H^2-v_C^2)\right]
\frac{\vec{v}_H + \vec{v}_C}{|\vec{v}_H+\vec{v}_C|^2}
\label{eq:w}
\end{equation}
Notice that $X$ in Eq.(\ref{eq:X}) has a non-negative denominator that becomes zero in an accidental event
where $\vec{v}_H$ and $\vec{v}_C$ are parallel;  $X$ is then ill-defined,
because no solution exists.  An alternate pair of $C$ and $H$ atoms must
be chosen.  In the simulations reported in subsequent sections, we find that $\Delta$ should be chosen small,
making the numerator of $X$ in Eq.(\ref{eq:X}) negative.  Thus both $X$ and $\alpha$ are negative, contrary
to the version shown in Fig. \ref{fig:spheres}.  This does not adversely affect anything.

There is a second solution, $\alpha = 1+\sqrt{1-X} $,
corresponding to the maximum $|\vec{w}|$, designated as $C^\prime$ in Fig. \ref{fig:spheres}.
For $|\vec{b}|>b_{\rm max}$, there are no real solutions.  This corresponds to $X>1$.
The condition for the two solutions to coincide is $X=1$, which agrees with $b_{\rm max}
=r(1+\cos\theta)$, where $\theta$ is the angle between $\vec{b}$ and $\vec{r}$.  This can be 
understood from the right hand part of Fig. \ref{fig:spheres}, illustrating the case where
the circle collapses to a point.  For reasonable choices of the parameter $m\delta/2=\Delta$,
meaning values smaller than or similar to $k_B T/N_S$, solutions should always exist.

\section{Debye-RTA model, especially $p$=2}
\label{app:A}

Appendix A gives formulas in the Debye-RTA model.
Here are details.  Equation 13 of ref. \onlinecite{Allen2014} is
\begin{equation}
\kappa_{\rm RTA}(q)=\frac{k_B}{\Omega}\sum_Q \frac{v_{Qx}^2}{1/\tau_Q+iqv_{Qx}}.
\label{eq:A1}
\end{equation}
This is the classical limit of the solution of the Boltzmann equation in the
relaxation time approximation (RTA).  Now make a Debye model, and model the
scattering rate as $1/\tau_Q=(1/\tau_D)(Q/Q_D)^p$.  Multiplying numerator and
denominator by $\tau_D$, and using $x=Q/Q_D$, $\mu=v_x/v=\cos\theta$, and $\lambda=qv$,
this generates Eq.(A1),
\begin{equation}
\tilde{\kappa}_p(q)=\frac{9}{2}\kappa_0 \int_0^1 dx x^2 \int_{-1}^1 d\mu \frac{\mu^2}{x^p + i\lambda\mu},
\label{eq:kp}
\end{equation}
where $\kappa_0=Nk_B v^2\tau/\Omega$.
One way to perform the integrations is to do the $\mu$-integral first.
\begin{equation}
\frac{\tilde{\kappa}_p(q)}{\kappa_0}=\frac{9}{\lambda} \int_0^1 dx x^2
\left[ \frac{x^p}{\lambda}-\left(\frac{x^p}{\lambda}\right)^2 \tan^{-1}\left(\frac{\lambda}{x^p}\right) \right] .
\label{eq:kp1}
\end{equation}

The most important case is $p=2$, and also the trickiest to integrate further.
Using $u=x^2$, this can be written as 
\begin{equation}
\frac{\tilde{\kappa}_2(q)}{\kappa_0}
= \frac{9}{2\lambda^2}\int_0^1 du u^{3/2}
\left[ 1-\frac{u}{\lambda} \cot^{-1}\frac{u}{\lambda} \right].
\label{eq:k2}
\end{equation}
This uses $\tan^{-1}(1/x)=\cot^{-1}x$.
The answer can be written as
\begin{equation}
\frac{\tilde{\kappa}_p(q)}{\kappa_0}=\frac{9}{7\lambda^2} [h(1)-h(0)]
\label{eq:A6}
\end{equation}
where $h(u)$ is
\begin{eqnarray}
&&h(u)=\frac{7}{2} \int du \left[ u^{3/2} - \frac{u^{5/2}}{\lambda} \cot^{-1}\frac{u}{\lambda}\right] \nonumber \\
&&=u^{5/2} + 2\lambda^2 u^{1/2} - \frac{u^{7/2}}{\lambda}\cot^{-1}\frac{u}{\lambda}
\nonumber \\
&&-\frac{\lambda^{5/2}}{\sqrt{2}}   \left[ \frac{1}{2}\log
\left( \frac{u+\sqrt{2u\lambda}+\lambda}{u-\sqrt{2u\lambda}+\lambda}\right) \right. \nonumber \\
&&\left. +\tan^{-1}\left(\sqrt{2u/\lambda} +1\right) +\tan^{-1}\left(\sqrt{2u/\lambda} -1\right) \right] 
\label{eq:A5}
\end{eqnarray}
This gives the result of Eq.(A4).  

Going back to Eq.(\ref{eq:kp}), an alternate route that is often simpler is to do the
radial ($x$) integral before the angular ($\mu$) integral.  In the tricky $p=2$ case, this gives
\begin{equation}
\frac{\tilde{\kappa}_2(q)}{\kappa_0}=9 \int_0^1 d\mu \mu^2
\left[ 1-\frac{\pi}{2}\sqrt{\frac{\lambda\mu}{2}} +{\cal R}{\rm e}\frac{r}{2}\log\left(\frac{1-r}{1+r}\right) \right] ,
\label{eq:A6}
\end{equation}
where $r=\sqrt{\lambda\mu}\exp(-i\pi/4)$.  By numerical integrations, we have convinced ourselves
that both integrals (Eq.(\ref{eq:kp1}) and Eq.(\ref{eq:A6})) give the same result as the fully
integrated formula in Eq.(A4).  These results are used in Appendix D to confirm the conjecture of 
Refs. \onlinecite{Allen2014} and \onlinecite{Sellan2010} 
that MD results, which are likely to conform to the $p=2$ case, should be extrapolated
by plotting $\kappa_{\rm eff}(L)$ {\it versus} $\sqrt(1/L)$.
%


%
%



\bibliographystyle{apsrev}
\bibliography{algori}